\documentclass[structabstract]{aa}  
\usepackage{hyperref}
\usepackage{graphicx}
\usepackage{natbib}
\usepackage{longtable}
\usepackage{caption}
\usepackage{subcaption}
\usepackage{txfonts}
\begin{document}
   \title{Infrared identification of high-mass X-ray binaries discovered by \textit{INTEGRAL}\thanks{Based on observations carried out at the European Southern Observatory (La Silla, Chile) under program IDs 080.D-0864(A) and 084.D-0535(A).}}

   \author{A. Coleiro \inst{1}, S. Chaty \inst{1,2},  J. A.  Zurita Heras\inst{3}, F. Rahoui\inst{4,5}, J. A. Tomsick\inst{6} }

   \institute{Laboratoire AIM (UMR-E 9005 CEA/DSM - CNRS - Universit\'e Paris Diderot), Irfu / Service d'Astrophysique, CEA-Saclay, 91191 Gif-sur-Yvette Cedex, France \email{alexis.coleiro@cea.fr}
 \and Institut Universitaire de France, 103 Boulevard Saint Michel, 75006 Paris, France \and  Fran\c{c}ois Arago Centre, APC, Universit\'e Paris Diderot, CNRS/IN2P3, CEA/Irfu, Observatoire de Paris, Sorbonne Paris Cit\'e, 10 rue Alice Domon et L\'eonie Duquet, 75205 Paris Cedex 13, France \and European Southern Observatory, Karl Schwarzschild-Strasse 2, 85748 Garching bei M\"unchen, Germany \and Department of Astronomy, Harvard University, 60 Garden Street, Cambridge, MA 02138, USA \and Space Science Laboratory, 7 Gauss Way, University of California, Berkeley, CA 94720-7450, USA}

   \date{Received / Accepted 09/13/2013}

  \abstract
 {Since it started observing the sky, the \textit{\textit{\textit{INTEGRAL}}} satellite has discovered new categories of high mass X-ray binaries (HMXB) in our Galaxy. These observations raise important questions on the formation and evolution of these rare and short-lived objects.}
   {We present here new infrared observations from which to reveal or constrain the nature of 15 \textit{\textit{INTEGRAL}} sources, which allow us to update and discuss the Galactic HMXB population statistics.}
   {After previous photometric and spectroscopic observing campaigns in the optical and near-infrared, new photometry and spectroscopy was performed  in the near-infrared with the SofI instrument on the ESO/NTT telescope in 2008 and 2010 on a sample of \textit{\textit{\textit{INTEGRAL}}} sources. These observations, and specifically the detection of certain features in the spectra, allow the identification of these high-energy objects by comparison with published nIR spectral atlases of O and B stars.}
   {We present photometric data of nine sources (IGR J10101-5654, IGR J11187-5438, IGR J11435-6109, IGR J14331-6112, IGR J16328-4726, IGR J17200-3116, IGR J17354-3255, IGR J17404-3655, and IGR J17586-2129) and spectroscopic observations of 13 sources (IGR J10101-5654, IGR J11435-6109, IGR J13020-6359, IGR J14331-6112, IGR J14488-5942, IGR J16195-4945, IGR J16318-4848, IGR J16320-4751, IGR J16328-4726, IGR J16418-4532, IGR J17354-3255, IGR J17404-3655, and IGR J17586-2129). Our spectroscopic measurements indicate that: five of these objects are Oe/Be high-mass X-ray binaries (BeHMXB), six are supergiant high-mass X-ray binaries (sgHMXB), and two are sgB[e]. From a statistical point of view, we estimate the proportion of confirmed sgHMXB to be 42\% and that of the confirmed BeHMXB to be 49\%. The remaining 9\% are peculiar HMXB.}
   %We also discuss the potential B[e] classification of IGR J10101-5654 that presents [\ion{Fe}{ii}] forbidden iron lines and H$_2$ emission line.}
   {}

\authorrunning{Coleiro et al.}
\titlerunning{Identification of HMXB discovered by \textit{INTEGRAL}.}
   \maketitle
%
%________________________________________________________________

\section{Introduction}
The \textit{\textit{INTEGRAL}} observatory has been observing the sky for ten years. By performing a detailed survey of the Galactic plane, it has discovered numerous new hard X-ray binary candidates that need to be identified. Of these, numerous high-mass X-ray binaries (HMXB) have been identified previously.\\
HMXB are interacting binary systems composed of a compact object orbiting around an O/B high-mass star. They are mainly located close to their formation sites in the Galactic plane \citep[see e.g.][]{Bodaghee2012,Coleiro2013}. These sources are typically separated into two principal families: Be X-ray binaries (BeHMXB) which consist of a neutron star that accretes matter when it moves through the equatorial decretion disk of a Be star, and supergiant X-ray binaries (sgHMXB), for which the donor object is an O/B supergiant star that feeds the compact object via its intense stellar wind. In this last category appear two subclasses of binary systems that were previously undetected. The first one consists in intrinsically highly obscured (with column densities N$_{\mathrm{H}}$ higher than about 10$^{23}$ cm$^{-2}$) hard X-ray sources, the most extreme example of which is the source IGR J16318-4848 \citep{Chaty2012}. The second one is composed of compact objects associated with a supergiant donor that undergo fast and transient outbursts in the X-ray domain of the electromagnetic spectrum. These latter sources are called supergiant fast X-ray transients (SFXT); \citep{Negueruela2006}, which represented by their archetype, IGR J17544-2619 \citep{Pellizza2006}.\\
Owing to their high intrinsic extinction and to interstellar absorption in the Galactic plane, detecting HMXB in longer wavelengths is challenging but crucial for any further study of their nature. In this context, near-infrared (nIR) observations appear to be very efficient in determining the nature of these HMXB by constraining the spectral type of the companion star.\\
In this article, we present results of an intensive nIR study of a sample of candidate HMXB for which an accurate X-ray localization is available. In Section \ref{obs} we describe the ESO nIR photometric and spectroscopic observations along with our data reduction strategy. In Section \ref{results} we review previously published properties of each source before reporting the results of our nIR observations. Finally, we discuss the results of the observations and implications in Section \ref{discuss} before concluding. 

\section{Observations and analysis} \label{obs}

The observations are based on astrometry, photometry, and spectroscopy of the 15 \textit{\textit{INTEGRAL}} sources listed in Table \ref{listHMXB}.
They were carried out on 2008 March 07-09 and 2010 March 27-29  at the European Southern Observatory (ESO, Chile), in nIR (1-2.5 $\mu$m) using the SofI instrument installed on the 3.5 m New Technology Telescope (NTT) at the La Silla observatory. These observations were performed through ESO programs ID  080.D-0864(A) and 084.D-0535 (PI Chaty), in visitor mode. 

\begin{table*}%AJOUTER A CE TABLEAU LES COORDONNEES DES CONTREPARTIES ETUDIƒES !!!!!!!!!!!!!
\begin{center}
\begin{tabular}{llllllcc}
\hline
  \hline
  Source & RA & Dec. & Unc. & $l$ & $b$ & Photometry & Spectroscopy  \\
  \hline
%  IGR J06074+2205 & 06:07:26.62 & +22:05:47.6 & 0.64" (Chandra)\\
%  IGR J08390-4833 & 08:38:48.9 & -48:31:25.3 & 2-2.5" (Chandra)\\
  IGR J10101-5654 & 10:10:11.87 & -56:55:32.1 & 0\farcs64 (Chandra) & 282.2568 & -00.6721& yes & yes \\
  IGR J11187-5438 & 11:18:21.1 & -54:37:32.0 & 3\farcs7 (Swift) & 289.6400   & 5.8260& yes & no  \\
%  IGR J11215-5952 & 11:21:46.9 & -59:51:46.9 & 1.1" (Swift)\\
  IGR J11435-6109 & 11:44:00.31 & -61:07:36.5 & 0\farcs64 (Chandra) & 294.8815   & 0.6865 & yes & yes \\
  IGR J13020-6359 & 13:01:59.2 & -63:58:06.0 & 3\farcs5 (Swift) & 304.0892 &    -1.1203 & yes & yes \\
  IGR J14331-6112 & 14:33:08.33 & -61:15:39.9 & 0\farcs64 (Chandra) & 314.8463  &  -0.7642 & yes & yes \\
  IGR J14488-5942  & 14:48:43.3 & -59:42:16.3 & 3\farcs7 (Swift) & 317.2340&   -0.1298 & yes & yes \\
  %IGR J14488-5942 b & 14:49:00.5 & -59:45:03.9 & 4.9" (Swift)\\
  IGR J16195-4945 & 16:19:32.20 & -49:44:30.7 & 0\farcs6 (Chandra) & 333.5570 &   0.3390 & yes & yes\\
  IGR J16318-4848 & 16:31:48.6 & -48:49:00 & 4\arcsec (XMM) & 335.6167  &  -0.4482 & yes & yes\\
  IGR J16320-4751 & 16:32:01.9 & -47:52:27 & 3\arcsec (XMM) & 336.3300  &  0.1689 & yes & yes\\
  IGR J16328-4726 & 16:32:37.88 & -47:23:42.3 & 1\farcs7 (Swift) & 336.7491   & 0.4220 & yes & yes\\
  IGR J16418- 4532 & 16:41:50.65 & -45:32:27.3 & 1\farcs9 (Swift) & 339.1883    & 0.4889 & yes & yes\\
   IGR J17200-3116 & 17:20:05.92 & -31:16:59.4 & 0\farcs64 (Chandra) & 355.0222   &  3.3474 & yes & no \\
%  IGR J17353-3539 & 17:35:23.5 & -35:40:13.8 & 3.5" (Swift)\\
  %IGR J17354-3255 a & 17:35:18.73 & -32:54:28.7 & 0.64" (Chandra)\\
  IGR J17354-3255  & 17:35:27.59 & -32:55:54.4 & 0\farcs64 (Chandra) &  355.4576  &  -0.2730 &yes & yes \\ 
  IGR J17404-3655 & 17:40:26.86 & -36:55:37.4 & 0\farcs64 (Chandra) & 352.6259 &   -3.2725 & yes & yes\\
  IGR J17586-2129 & 17:58:34.56 & -21:23:21.6 & 0\farcs64 (Chandra) & 7.9862   & 1.3266 & yes & yes\\
  \hline
\end{tabular}
  \caption{X-ray position, uncertainty for the observed X-ray sources, and galactic longitude ($l$), and latitude ($b$). Observing mode(s) (photometry and/or spectroscopy) are also given for each source.}
    \label{listHMXB}
\end{center}
\end{table*}

\begin{table*}%ENLEVER LES COORDONNEES ICI
\begin{tabular}{lllllllll}
\hline
  \hline
  Source & RA & Dec. & Obs. Date & AM & ET & $J$ mag. & $H$ mag. & $K_s$ mag.\\
  \hline
    \multicolumn{9}{c}{\textbf{2008 Run}}\\
IGR J11187-5438  & 11:18:21.21 & -54:37:28.6 & 2008-03-09T07:08:39.930 & 1.2 & 10.0 & 15.66 $\pm$ 0.04 & 14.99 $\pm$ 0.02 & 14.68 $\pm$ 0.02\\
 IGR J11435-6109 & 11:44:00.3 & -61:07:36.5 & 2008-03-09T05:54:12.202 &  1.2 & 10.0 &12.90 $\pm$ 0.02 & 12.23 $\pm$ 0.02 & 11.77 $\pm$ 0.02\\
 IGR J17200-3116 & 17:20:06.1 & -31:17:02.0 & 2008-03-09T08:35:06.750 & 1.1 & 10.0 & 13.50 $\pm$ 0.02 & 12.48 $\pm$ 0.02 & 12.04 $\pm$ 0.02\\
 
    \multicolumn{9}{c}{\textbf{2010 Run}}\\

%IGR J08390-4833 & 08:38:48.9 & -48:31:25.2 & 2010-03-29T00:22:26.433 & 1.1 & 10.0 & -- & -- & 17.287\\
 IGR J10101-5654 & 10:10:11.87 & -56:55:32.1 & 2010-03-29T01:17:11.318 & 1.2  & 2.0 & -- & -- & 10.742 $\pm$ 0.02\\
 % IGR J13020-6359 a & 13:01:59.2 & -63:58:05.9  & 2010-03-29T03:16:57.383 & 1.3 & 10.0 & -- & -- & 17.246 ???\\
 %IGR J13020-6359 b &  &   &  & & & -- & -- & 16.534\\
 %IGR J13020-6359 c &  &   &  & & & -- & -- & 16.633\\
 IGR J14331-6112 & 14:33:08.33 & -61:15:39.7 & 2010-03-28T04:53:33.431 & 1.3 & 10.0 & -- & -- & 13.691 $\pm$ 0.02\\
 IGR J16328-4726 & 16:32:37.91 & -47:23:40.9 & 2010-03-29T07:45:12.428 & 1.1 & 10.0 & -- & -- & 11.309 $\pm$ 0.02\\ %saturated 20K ADU...
IGR J17354-3255 & 17:35:27.60 & -32:55:54.40 & 2010-03-30T06:51:03.000 & 1.3 & 10.0 & -- & -- & 10.395 $\pm$ 0.02\\
IGR J17404-3655 & 17:40:26.85  & -36:55:37.6 & 2010-03-30T08:10:58.684 & 1.1 & 10.0 & -- & -- & 14.370 $\pm$ 0.02\\
IGR J17586-2129 & 17:58:34.56 & -21:23:21.53 & 2010-03-30T07:44:31.981 & 1.2 & 2.0 & -- & -- & 9.4\\ %saturated 18K ADU...
\hline
\end{tabular}
  \caption{Counterpart position and  photometry results from this study. We point out that the $K_s$-band magnitude of IGR J17586-2129 is not accurate since photometric data of this source were saturated. Airmass (AM) and single exposure time in seconds (ET) are also given for each source.}
    \label{list_counterparts}
\end{table*}

\begin{table*}
\begin{center}
\begin{tabular}{clllll}
  \hline
  \hline
  Source  & Obs. Date & AM & ET & Filter & Slit Width\\
  \hline
 % \multicolumn{6}{c}{\textbf{Run 2008}}\\
 %IGR J06074+2205  & 2008-03-09T00:58:44.5197 & 1.7 & 60.0 & GRF &1.0 \\
 %IGR J11215-5952 &2008-03-09T05:05:55.6092 &1.6 & 60.0 & GRF &1.0 \\
 %IGR J11435-6109 &2008-03-09T06:32:30.0682 &1.2 & 60.0 & GRF & 1.0\\
  % \multicolumn{6}{c}{\textbf{Run 2010}}\\
% IGR J08390-4833 & 2010-03-29T00:38:17.5258 & 1.1 & 60.0 & $K_s$ & 1.0\\
 IGR J10101-5654 &  2010-03-28T00:21:11.7584 & 1.2 & 60.0 & $K_s$ & 0\farcs6\\
 " & 2010-03-30T00:22:04.8787 & 1.2 & 60.0 & H &1 \farcs0\\
 IGR J11435-6109 & 2010-03-29T02:37:37.6419 & 1.2 & 60.0 & $K_s$ & 0 \farcs6\\
 " & 2010-03-30T02:37:54.2772 & 1.2 & 60.0 & H & 1\farcs0\\
 IGR J13020-6359 & 2010-03-29T03:35:41.7071 & 1.3 & 60.0 & $K_s$ & 1\farcs0\\
 IGR J14331-6112 & 2010-03-28T05:08:59.2288 & 1.2 & 60.0 & $K_s$ & 1\farcs0\\
 IGR J14488-5942 & 2010-03-29T05:56:02.3304 & 1.2 & 60.0 & $K_s$ & 1\farcs0\\
 IGR J16195-4945 & 2010-03-30T04:42:57.4456 & 1.5 & 60.0 & $K_s$ & 1\farcs0\\
 " & 2010-03-30T03:59:32.1327 & 1.7 & 60.0 & H & 1\farcs0\\
 IGR J16318-4848 & 2010-03-28T05:51:33.3738 & 1.3 & 60.0 & $K_s$ & 0\farcs6\\
 IGR J16320-4751 &  2010-03-28T06:37:01.7617 & 1.2 & 60.0&$K_s$ & 1\farcs0\\
 IGR J16328-4726 & 2010-03-29T08:03:29.7491 & 1.1 & 60.0 & $K_s$ & 1\farcs0\\
 IGR J16418-4751 & 2010-03-28T07:31:53.0447 & 1.1 & 60.0&$K_s$ & 1\farcs0\\
% IGR J17353-3539 & 2010-03-30T05:28:07.9150 & 1.7 & 60.0 & $K_s$ & 1.0\\
 IGR J17354-3255 & 2010-03-30T07:05:38.8785 & 1.2 & 60.0 & $K_s$ & 1\farcs0\\
 IGR J17404-3655 & 2010-03-30T08:31:51.5539 & 1.1 & 60.0 & $K_s$ & 1\farcs0\\
 IGR J17586-2129 & 2010-03-30T07:56:13.1784 & 1.2 & 5.0  & $K_s$ & 1\farcs0\\

    \hline
\end{tabular}
  \caption{Log of nIR spectra. The name of the source, the date and UT time of the observations, the airmass (AM), the exposure time in seconds (ET), the filter and the slit width are indicated. All spectra were obtained at ESO/NTT with SofI instrument.}
    \label{log_spectra}
\end{center}
\end{table*}

\subsection{Photometry}

We performed nIR photometry in the $J$-, $H$-, and $K_s$-bands of the sources listed in Table \ref{list_counterparts}.
We used the large field of SofI's detector, giving an image scale of 0\farcs288/pixel and a field of view of 4\farcm92 $\times$ 4\farcm92. The photometric observations were obtained by repeating a set of images for each filter with nine different 30\arcsec offset positions including the targets, following the standard jitter procedure, which enable us to cleanly subtract the sky emission in nIR. Each individual frame has an integration time of 10 s, giving a total exposure time of 90 s in each energy band. 
Moreover, three photometric standard stars chosen in Persson's catalog \citep{Persson1998} were observed in the three bands with a total integration time of 10 s for each target in each band.\\

We used the IRAF (Image Reduction and Analysis Facility) suite\footnote{IRAF is distributed by the National Optical Astronomy Observatories, which are operated by the Association of Universities for Researc in Astronomy, Inc., under cooperative agreement with the National Science Foundation.} to perform data reduction and carry out standard procedures of nIR image reduction, including flat-fielding and nIR sky subtraction.\\

We performed accurate astrometry on each entire SofI 4\farcm92 $\times$ 4\farcm92 field, using all stars from the 2MASS catalog in this field. The rms of the astrometry fit was always lower than 0\farcs6.\\ %The finding charts including the results of our astrometry are shown in Fig. ?? for all sources of our sample.\\
We carried out aperture photometry on photometric standard stars and computed the zero-point value for each energy band knowing the instrumental magnitudes and the apparent magnitudes (given in Persson's catalog) and using the standard relation $mag_{app}=mag_{inst}-Zp$, where $mag_{app}$ and $mag_{inst}$ are the apparent and instrumental magnitudes, respectively, and $Zp$ is the zero-point, including the extinction and airmass terms. %??
After computing the zero-point value for each filter, we performed PSF-fitting photometry on the crowded fields that contained the six sources under study following the standard utilization of the IRAF \textit{noao.digiphot.daophot} package. After evaluating an aperture correction for each filter and each object (due to the difference of aperture radius used to perform photometry on standard stars and targets) and knowing the zero-point values, we derived the apparent magnitude of the targets in the three nIR filters. The results are given in Table \ref{list_counterparts}.\\

\subsection{Spectroscopy}

We also carried out nIR spectroscopy with SofI. In 2008, low-resolution spectra were obtained between 1.53 and 2.52 $\mu$m, but, spectra are not presented in this paper because of the low signal-to-noise ratio (S/N) that was caused by bad weather which prevented us from extracting any useful data. In 2010, we used the medium-resolution grism with $H$ and $K_s$ filters (properties of the grisms are given in Table \ref{spectro_filters}). For each source, eight spectra were taken with both filters, half of them with the 1\farcs0 (or 0\farcs6) slit on the source and the other half with an offset of 30\arcsec, in order to subtract the nIR sky emission. Each individual spectrum has an exposure time of 60 s, giving a total integration time of 480 s. Furthermore, four telluric standards were observed immediately after each target with the same instrument set-up and a total integration time of 8 s. A summary of the spectroscopic observations is provided in Table \ref{log_spectra}.\\

%+spectres moyenne rŽsolution (2010)

\begin{table*}
\begin{center}
\begin{tabular}{llllll}
  \hline
  \hline
  Filter name & Wavelength range (microns) & Slit width & Resolving power  & Dispersion (\AA/pixel)\\
  \hline
 % GRF & 1.53-2.52 & 588 -- 980 & 10.22\\
  $H$ & 1.50-1.80 & 1\farcs0 & 900  & 3.43\\
    $H$ & 1.50-1.80 & 0\farcs6 & 1500 & 3.43\\
   $K_s$ & 2.00-2.30 & 1\farcs0 & 1320  & 4.62\\
   $K_s$ & 2.00-2.30 & 0\farcs6 &  2200  & 4.62\\
   \hline
\end{tabular}
  \caption{Wavelength range and resolution for each grism. The resolution is given for the 1\farcs0 and 0\farcs6 slit respectively.}
    \label{spectro_filters}
\end{center}
\end{table*}

We analyzed the nIR spectra using standard IRAF tasks, correcting for flat field, removing the crosstalk, correcting the geometrical distortion, combining the images and finally extracting the spectra and performing wavelength calibration using the IRAF \textit{noao.twodspec} package. Wavelength calibration was made with a Xenon lamp which presents a good distribution of lines and is sufficient to calibrate data taken with the low-resolution grisms of the SofI instrument. The target spectra were then corrected for the telluric lines using the standard stars observed with the same configuration. 

%\section{Data reduction}\label{datareduc}

\section{Results}\label{results}

All sources studied in this paper were discovered with the IBIS/ISGRI detector onboard the \textit{\textit{\textit{INTEGRAL}}} observatory. Furthermore, they were all observed with other X-ray facilities to provide an accurate localization, which allowed us to determine the optical/nIR counterpart. The sample of 15 sources is given in Table \ref{listHMXB}. We present below our results on each source, for which we followed the same strategy. We first observed the field in nIR, performed accurate astrometry, and derived the photometry of the counterpart candidate for most of them. We then analyzed the nIR spectrum.\\
We also underline that the low S/N and the medium-resolution of the spectra prevent a quantitative study. Moreover, important lines such as \ion{He}{i} at 2.058 $\mu$m and Br$\gamma$ at 2.1661 $\mu$m are often subject to difficulties because of the strong telluric absorption present at these wavelengths. Thus, considering these issues, we mostly conducted a qualitative analysis based on a comparison of our nIR spectra with available nIR spectral atlases \citep{Hanson1996, Hanson2005, Morris1996, McGregor1988, Clark1999, Steele2001}. In general, we considered as a stellar line (and not a spurious feature) each line detected at least at 3$\sigma$ of the noise estimated locally, that had a width broader than the instrumental width of 20 $\AA$ for the H filter and 18 $\AA$ for the $K_s$ filter with the 1\farcs0 slit.\\
Each detected line was fitted by a Gaussian profile using the IRAF \textit{splot} task. We list its fitted position $\lambda_{fit}$, its equivalent width (EW), its full width at half maximum (FWHM), and its flux in Tables \ref{lines_all} and \ref{lines_all_2}. We estimate the average error on these parameters to be 20\% because of the noise and localization of the continuum. Finally, according to the resolution of the instrument, we considered that a feature can be offset from its laboratory rest wavelength. This offset $\Delta\lambda$ (computed for the 1\farcs0 slit and approximated by $\Delta\lambda = \lambda/R$, where R is the resolving power given in Table \ref{spectro_filters}) can be as large as 20 $\AA$ for the H filter and 18 $\AA$ for the $K_s$ filter.

%EVOQUER LES INCERTITUDES SUR LES FWHM ET EW
%SPECTRA OF ... ARE not shown because T\ion{He}{i}R faintness prevented us from extracting useful data.

%-----------------IGR J10101-5654------------------------------------------------------------------------------------

\subsection{IGR J10101-5654}\label{10101_biblio}

IGR J10101-5654 was discovered on 2006 January by \citet{Kuiper2006} with the instrument IBIS ISGRI at the position RA = 10$^{\rm{h}}$10$^{\rm{m}}$07$^{\rm{s}}$.8, Dec = -56$^\circ$54\arcmin46\farcs4 (equinox J2000.0; uncertainty 2\farcm2). \citet{Masetti2006} localized an optical counterpart at RA = 10$^{\rm{h}}$10$^{\rm{m}}$11$^{\rm{s}}$.866, Dec = -56$^\circ$55\arcmin32\farcs06 (equinox J2000.0; accuracy better than 0\farcs1) and the optical spectrum is typical of an HMXB optical counterpart, showing strong narrow H$\alpha$ emission superimposed on a reddened continuum. The lack of reliable optical photometry for this counterpart prevented them from deriving a spectral type for this object. Nevertheless,  the H$\alpha$ EW value appears too high for a supergiant star. Thus, the authors suggest that this HMXB hosts a secondary star of intermediate luminosity class (early giant). \citet{Tomsick2008} confirmed the counterpart suggested by \citet{Masetti2006}. Their \textit{Chandra} spectrum also gives a column density of N$_H$ = 3.2$^{+1.2}_{-1.0}\times$10$^{22}$ cm$^{-2}$ and does not indicate a high level of local absorption. The \textit{Chandra} flux, extrapolated into the 20--40 keV  \textit{\textit{\textit{INTEGRAL}}} band, appears to be significantly lower than the flux measured by \textit{\textit{\textit{INTEGRAL}}} (flux ratio of 0.05 $\pm$ 0.04). This indicates an important change in mass accretion rate onto the compact object and could be explained by an eccentric binary orbit \citep{Tomsick2008}. \textit{Swift} observations analyzed by \citet{Rodriguez2009} finally confirmed the previously suggested association for this object, that it is very likely an HMXB and also the possible substantial variation of the mass accretion rate.\\

NIR spectra of the source are shown in Figure \ref{first}. We report the detected lines in Table \ref{lines_all}. The $H$-band spectrum exhibits the Brackett series from Br(20-4) line at 1.5198 $\mu$m to the Br(10-4) transition at 1.7377 $\mu$m. Two lines could match the position of $\left[\rm{\ion{Fe}{ii}}\right]$ lines at 1.6427 $\mu$m and 1.7122 $\mu$m (this hypothesis is discussed at the end of the paragraph). The $K_s$-band spectrum clearly shows \ion{He}{i} at 2.0586 $\mu$m and Br(7-4) at 2.1659 $\mu$m. \ion{Mg}{ii}, \ion{He}{ii}, \ion{Na}{i}, and \ion{Fe}{ii} emission lines are also detected in this spectrum. The nIR spectra are typical of a Be companion star and the intensity ratio of the Br(7-4) and \ion{He}{i} (at 2.0586 $\mu$m) lines suggests a B0.5Ve or a B0Ivpe type \citep{Hanson1996}. However, the $\left[\rm{\ion{Fe}{ii}}\right]$ emission lines are confusing. Indeed, these forbidden lines are typical of the B[e] phenomenon and are tracers of recombination in a dense stellar wind \citep{Clark1999}. The \ion{Na}{i} emission line suggests the presence of a low-excitation region that is not directly exposed to the radiation of the star or the compact object. It is probably the signature of an extended circumbinary envelope \citep{McGregor1988}. A feature at 2.1226 $\mu$m might be associated with the molecular hydrogen H$_2$ ro-vibrational $v$ = 1-0, J = 3-1 S(1) emission line. If this feature is real, its presence and that of $\left[\rm{\ion{Fe}{ii}}\right]$ lines could trace supersonic shocks that heat the gas \citep[see e.g.][]{Clark1999, Chen1998}. These processes are very similar to those generally observed in the close environment of young stellar objects (YSOs) \citep[see e.g.][]{Chen1998}. Finally, the \ion{He}{i} line at 2.058 $\mu$m shows a double-peaked shape. The separation is $c\frac{\Delta\lambda}{\lambda}=290$ km s$^{-1}$, larger than the resolution of the instrument, $c\frac{\Delta\lambda}{\lambda}\sim260$ km s$^{-1}$. While this double-peak structure could be the signature of a circumstellar disk, this result remains uncertain. Then we suggest that this source is surrounded by a highly stratified environment composed of regions shielded from the direct stellar radiation.\\
Moreover, we fitted the SED from optical to mIR wavelengths \footnote{We note that B and R magnitudes, coming from the USNO survey seem to be unreliable. Because of the faintness of the source in the optical, we did not take optical magnitudes into account in the fitting procedure.} with a model that combines two absorbed blackbodies (see Figure \ref{best_fit}), one representing the companion star emission and another one representing a possible mIR excess due to absorbing material that enshrouds the companion star \citep[see e.g.][]{Rahoui2008}. The free parameters of the fits are the absorption in V band, $A_V$, the radius-to-distance ratio $R_*/D$, the additional blackbody component temperature $T_d$, and its radius $R_d$. The stellar blackbody temperature is fixed to 20000 K. The best-fitting parameters are $A_V$ =13.1 $\pm$ 0.86 mag, $R/D$ = 2.79 $\pm$ 0.20 $R_{\sun}$/kpc,  $T_d$ = 1016 K $\pm$ 127, $R_d/D$ = 28.3 $\pm$ 4.25 $R_{\sun}$/kpc, and $\chi^2/dof = 2.23$. For comparison, the best-fitting parameters without the additional component are $A_V$ =17.01 $\pm$ 0.46 mag, $R/D$ = 3.86 $\pm$ 0.09 $R_{\sun}$/kpc, and $\chi^2/dof = 23.58$.\\

 To conclude, SED fitting shows a clear mIR excess, most likely caused by the presence of warm dust around the system. Moreover, nIR spectra of IGR J10101-5654 present many similarities with stars that show the B[e] phenomenon, even though the optical spectrum of this source, published by \citet{Masetti2006}, appears to be very different from the optical spectrum expected for a B[e] phenomenon. Indeed, we can reasonably assume that in the optical spectrum, one mainly detects radiation from the stellar photosphere whereas the nIR spectra mostly probes the circumstellar environment, which is characterized by a B[e] phenomenon. We suggest that IGR J10101-5654 is a supergiant B[e]. Figure \ref{comparison} compares the $K_s$ spectrum of the source with $K_s$ spectra of IGR J11435-6109 and IGR J13020-6359. These spectra seem very similar but the Br$\gamma$ emission line of IGR J10101-5654 looks broader than those of the last two. 
%
%\begin{figure}[h!]
%\begin{center}
%\includegraphics[scale=0.45]{fit_10101.pdf}
%\caption{Best fit of IGR J10101-5654. Derived parameters: $A_V$ =13.1 mag; $T_*$ = 20000 K; $R/D$ = 2.79 $R_{\sun}$/kpc;  $T_d$ = 1016 K; $R_d$ = 28.3 $R_{\sun}$/kpc; $\chi^2/dof = 2.23$.}
%\label{fit10101}
%\end{center}
%\end{figure}

%-----------------IGR J11187-5438---------------------------------------------------------------------------------------

\subsection{IGR J11187-5438}
IGR J11187-5438 was discovered by \citet{Bird2007} at the position RA = 11$^{\rm{h}}$18$^{\rm{m}}$21$^{\rm{s}}$.1, Dec = -54$^\circ$37\arcmin32\arcsec (equinox J2000.0; uncertainty 4\arcmin). \citet{Rodriguez2008} conducting \textit{Swift} observations refined the position of the source and found a single source in the 2MASS catalog (2MASS J11182121-5437286) that is also visible in the DSS II catalog. The 0.5-9 keV spectrum is well fitted by an absorbed power law with $\Gamma = 1.5$ and N$_H$ = 0.28$\times$10$^{22}$ cm$^{-2}$ similar to the Galactic absorption along the line of sight meaning that the source is not intrinsically absorbed.\\

We fitted the SED, constituted by the $I$, $J$, $H$, $K_s$, WISE 3.4, and WISE 4.6 band magnitudes (see Table \ref{WISE}), with a stellar blackbody model with three free parameters: the companion star temperature $T$, the ratio radius of the companion star over the distance, $R/D$, and the extinction in the V band $A_V$. The retrieved parameters are not consistent with early star parameter values: $A_V = 3.9 \pm 0.4$ mag, T = 6000 K, R/D = 0.60 $\pm$ 0.03 $R_{\sun}$/kpc, and $\chi^2/dof = 3.45$ (see Figure \ref{best_fit}). Thus, this source is probably not an HMXB but it might be a low mass X-ray binary (LMXB).

%\begin{figure}[h!]
%\begin{center}
%\includegraphics[scale=0.35]{fit11187.pdf}
%\caption{Best fit of IGR J11187-5438. Derived parameters: $A_V = 3.5$ mag; T = 6000 K; R/D = 1.33 $\times$ 10$^{-11}$; $\chi^2_{min} = 4.1$.}
%\label{11187fit}
%\end{center}
%\end{figure}

%-----------------IGR J11435-6109------------------------------------------------------------------------------------

\subsection{IGR J11435-6109}

IGR J11435-6109 was discovered by \citet{Grebenev2004} at the position RA = 11$^h$43$^{\rm{m}}$52$^{\rm{s}}$, Dec = -61$^\circ$09\arcmin00\arcsec, (equinox J2000.0;  uncertainty 2\farcm5). The optical and infrared counterparts were discovered by \citet{Tomsick2007} and confirmed by \citet{Negueruela2007_1} as 2MASS J11440030-6107364 = USNO-B1.0 0288-0337502. \textit{BeppoSAX-WFC} observations carried out by \citet{int-Zand2004} and \citet{Swank2004} suggested a column density of N$_H$ $\sim$ 9$\times$10$^{22}$ cm$^{-2}$ and a $\Gamma$ value of $\sim$ 1.9 - 2.3. These values were confirmed by \citet{Tomsick2008} who fitted the {Chandra}/ACIS spectrum with a power-law and found an N$_H$ value of 15$\times$10$^{22}$ cm$^{-2}$ and a $\Gamma$ value of $\sim$ 1.1. \citet{int-Zand2004} detected a pulsation of 161.76$\pm$0.01s and a possible orbital period of 52.5 days which was confirmed by \citet{Corbet2005}  and \citet{Wen2006} with \textit{RXTE/ASM} observations that gave an orbital period of 52.46$\pm$0.06 days and 52.36 days respectively. \citet{Tomsick2007}, thanks to \textit{Chandra} observations, concluded that IGR J11435-6109 is an HMXB with a significant amount of intrinsic absorption. This result was confirmed by optical observations of the source, conducted by \cite{Negueruela2007}: the source is not detected below 5000 $\AA$ whereas the 5000-7800 $\AA$ spectrum shows a strong H$\alpha$ emission line on top of a reddened continuum. The equivalent width of the H$\alpha$ line is equal to -26 $\AA$, and the lack of other stellar features enabled them to classify the companion star as an obscured Be star, compatible with the position of the system in the Corbet diagram \citep{Corbet2005}. Finally, the spectral type was constrained by \citet{Masetti2009} who suggested a B2III or B0V counterpart with an extinction A$_{\rm{V}}$ of 5.7.\\

The nIR spectra of the source are shown in Figure \ref{first}. We report the detected lines in Table \ref{lines_all}. The $H$-band spectrum exhibits the Brackett series from the Br(21-4) line at 1.5137 $\mu$m to the Br(10-4) transition at 1.7364 $\mu$m. An additional line may match with the position of $\left[\rm{\ion{Fe}{ii}}\right]$ line at 1.7120 $\mu$m and another one, located at 1.7014 $\mu$m, may match with an \ion{He}{i} line. \ion{N}{iii}, \ion{C}{iii} and \ion{Fe}{ii} may also be detected around 1.5772 $\mu$m. The $K_s$ band spectrum clearly shows \ion{He}{i} at 2.0590 $\mu$m and Br(7-4) at 2.1664 $\mu$m. \ion{Mg}{ii} and \ion{He}{ii} emission lines are also present in this spectrum. The nIR spectra are pretty similar to those of IGR J10101-5654 but fewer lines are detected. Especially some typical lines of stars that exhibit B[e] phenomenon are not observed. Thus, we consider these spectra typical of a Be companion star and the intensity ratio of the Br(7-4) and \ion{He}{i} (at 2.0590 $\mu$m) lines suggests a B0.5Ve \citep{Hanson1996} (see comparison spectra in Figure \ref{comparison}, left panel). Moreover, we fitted the SED from optical to mIR wavelengths\footnote{We note that B and R magnitudes, coming from the USNO survey seem to be unreliable. Because of the faintness of the source in the optical, we did not take optical magnitudes into account in the fitting procedure.} with a model that combines two absorbed blackbodies (see Figure \ref{best_fit}), one representing the companion star emission and another one representing a possible mIR excess due to absorbing material enshrouding the companion star \citep[see e.g.][]{Rahoui2008}. The free parameters of the fits are the absorption in V band, $A_V$, the radius-to-distance ratio $R_*/D$, the additional blackbody component temperature $T_d$ and its radius $R_d$. The stellar blackbody temperature was fixed to 20000 K. The best-fitting parameters are $A_V$ =7.98 $\pm$ 0.44 mag,  $R/D$ = 1.33 $\pm$ 0.05 $R_{\sun}$/kpc, $T_d$ = 925.6 K $\pm$ 190.7, $R_d/D$ = 51.6 $\pm$ 2.97 $R_{\sun}$/kpc, $\chi^2/dof = 1.14$. For comparison, the best-fitting parameters without the additional component are $A_V$ =8.98 $\pm$ 0.18 mag, $R/D$ = 1.46 $\pm$ 0.03 $R_{\sun}$/kpc, and $\chi^2/dof = 6.81$. Thus SED fitting shows a relatively weak mIR excess, which we consider to be probably due to bremsstrahlung emission coming from the decretion disk of a Be star. Thus, we confirm the Be type of this HMXB and constrain its spectral type to B0.5Ve.

%-----------------IGR J13020-6359-------------------------------------------------------------------------------------

\subsection{IGR J13020-6359}

IGR J13020-6359 was discovered in 2006 by \citet{Bird2006} and was classified as a pulsar/HMXB in \citet{Bird2007} based on its position coinciding with 2RXP J130159.6--635806 studied by \citet{Chernyakova2005} who conducted multi-epoch \textit{ASCA}, \textit{Beppo-SAX}, \textit{XMM-Newton} and \textit{\textit{\textit{INTEGRAL}}} observations. The observed long term behaviour and the spectral and timing properties tend to indicate a HMXB with Be companion. These authors identified a probable 2MASS infrared counterpart at (RA, Dec J2000.0) = (13$^h$01$^m$58$^s$.7, -63$^\circ$58\arcmin09\arcsec). The \textit{Swift} spectrum, well fitted with a power law, shows some slight variability as expected for an accreting pulsar \citep{Rodriguez2009}.\\
 
%Žvoquer l'incertitude sur la contrepartie !

\begin{figure*}[h!]
\begin{minipage}[t]{1\textwidth}
\begin{center}
\includegraphics[width=0.5\textwidth]{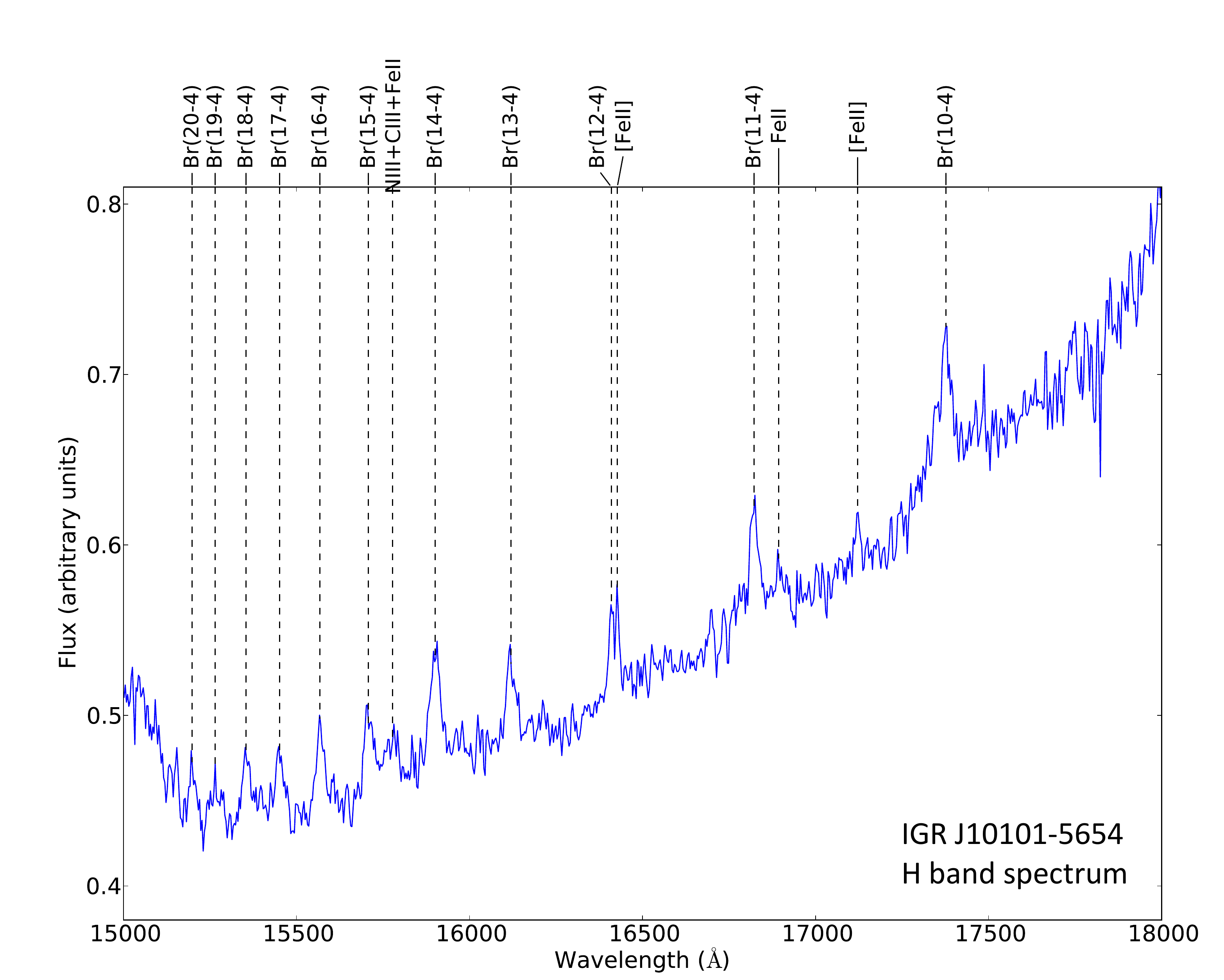}\hfill
\includegraphics[width=0.5\textwidth]{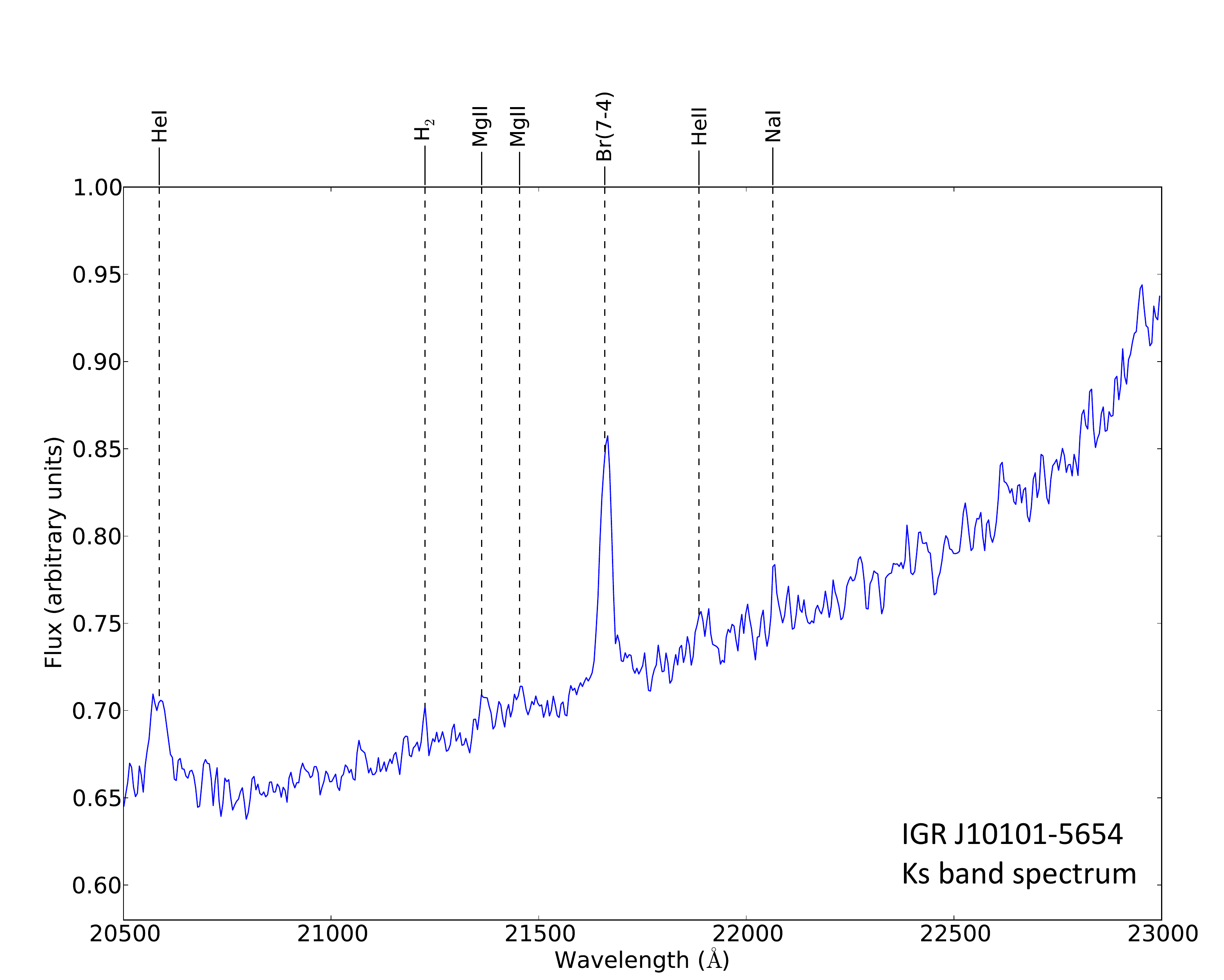}\\
\includegraphics[width=0.5\textwidth]{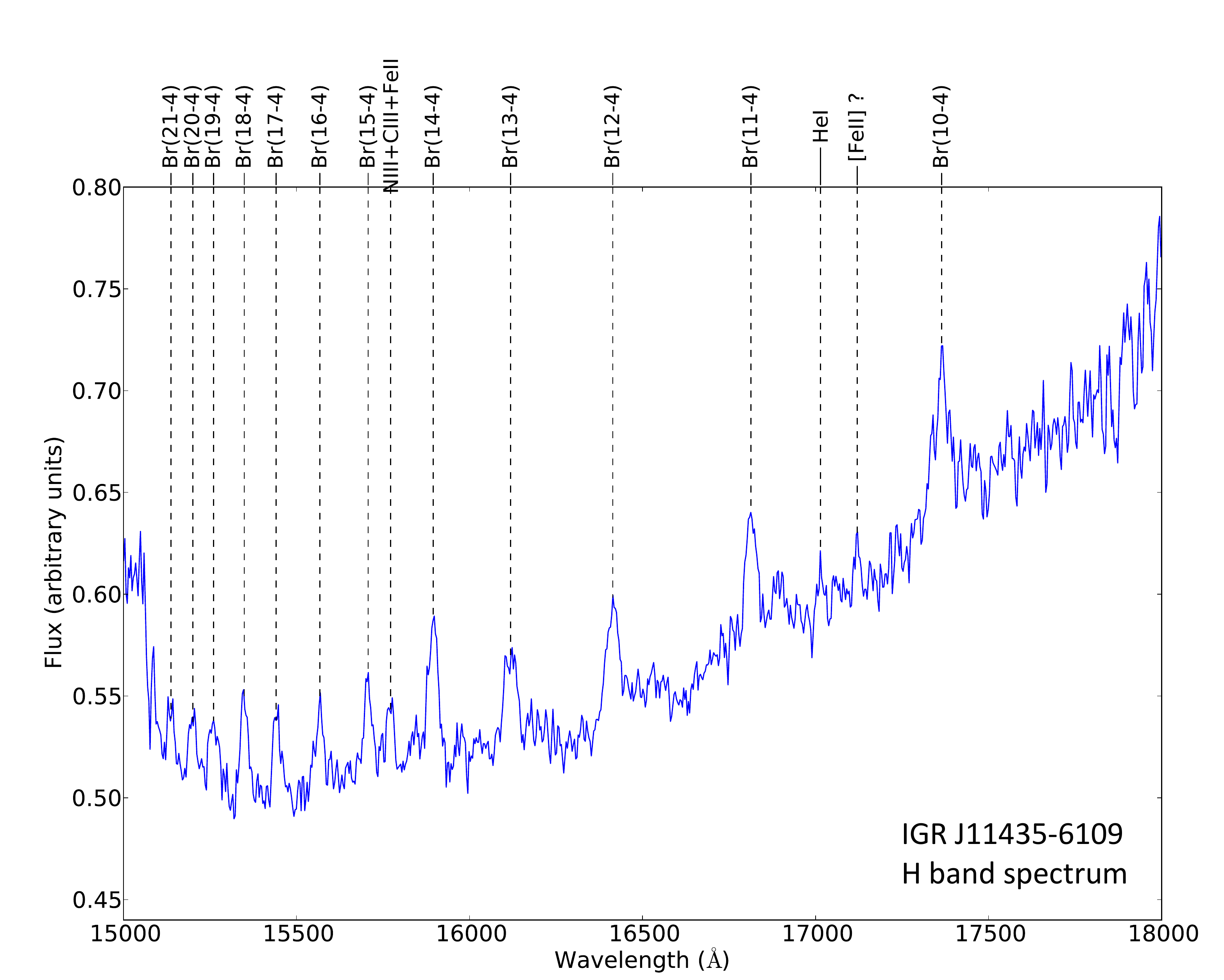}\hfill
\includegraphics[width=0.5\textwidth]{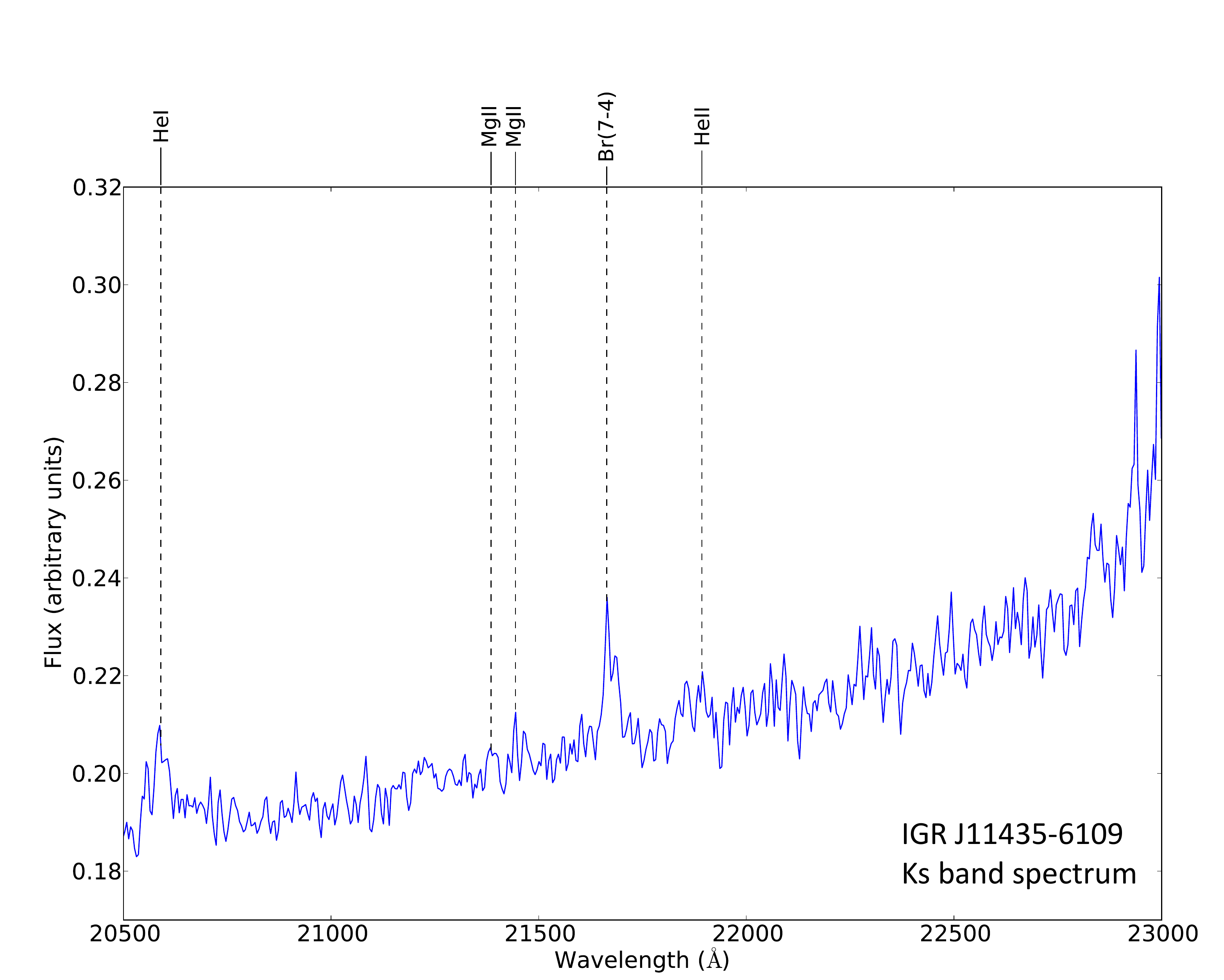} \\
\includegraphics[width=0.5\textwidth]{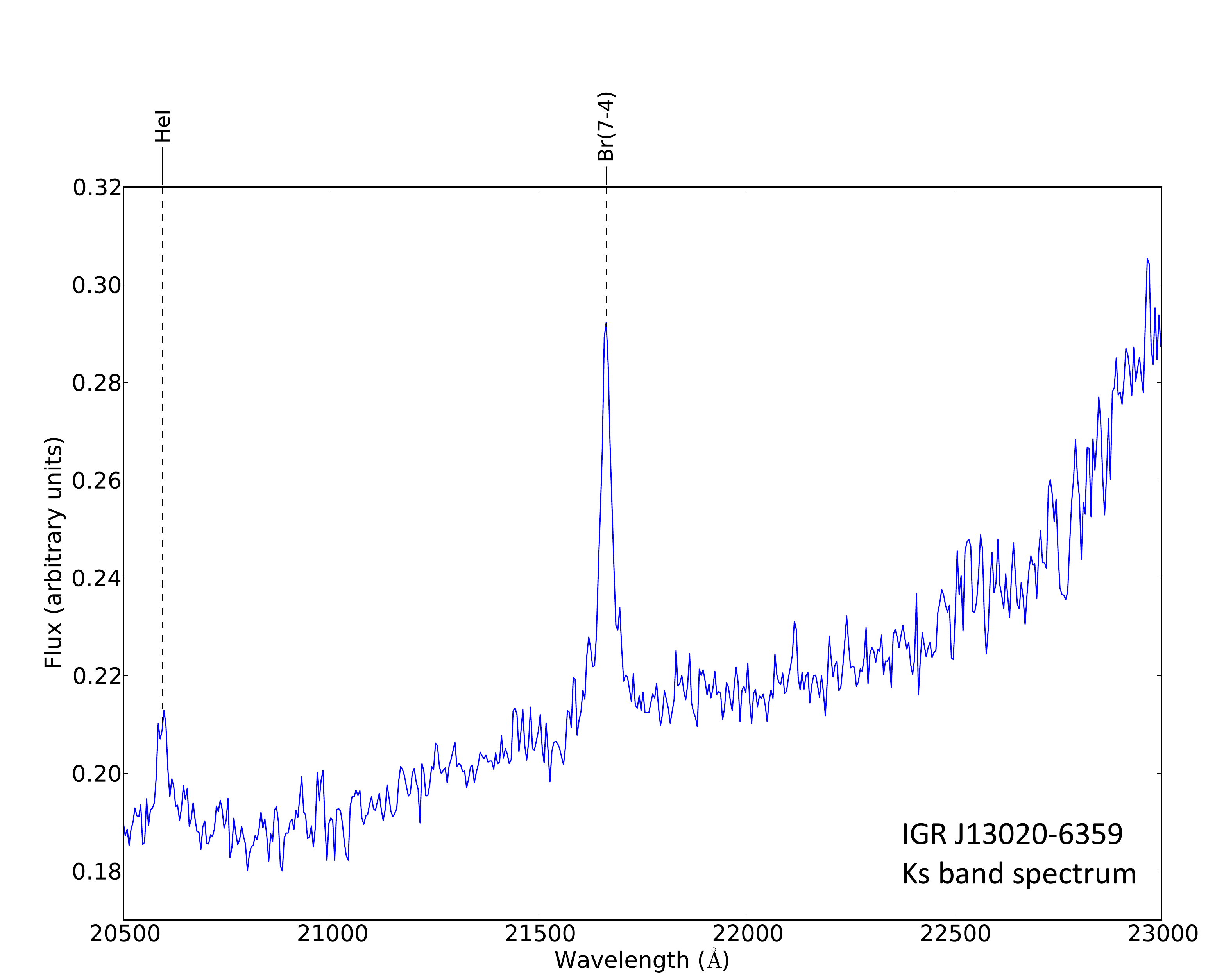}\hfill
\end{center}
\end{minipage}
\caption{Spectra of IGR J10101-5654 (H band: top left and $K_s$ band: top right), IGR J11435-6109 (H band: middle left and $K_s$ band: middle right), and IGR J13020-6359 (bottom right).}
\label{first}
\end{figure*}

The $K_s$-band spectrum of the source is shown in Figure \ref{first}. We report the detected lines in Table \ref{lines_all}. This spectrum clearly shows the \ion{He}{i} at 2.0594 $\mu$m and Br(7-4) at 2.1663 $\mu$m. Even though no other lines are detected, the nIR spectrum is typical of a Be companion star and the intensity ratio of the Br(7-4) and \ion{He}{i} (at 2.0594 $\mu$m) lines suggests a B0.5Ve or a B0Ivpe type \citep{Hanson1996}, similar to IGR J10101-5654 and IGR J11435-6109 (see Figure \ref{comparison}, left panel). We therefore confirm the Be type of this HMXB and constrain its spectral type to B0.5Ve.

%\begin{figure*}[th!]
%\begin{center}
%\includegraphics[scale=1]{firstgroup.pdf}
%\caption{Spectra of IGR J10101-5654, IGR J11435-6109, IGR J13020-6359.}
%\label{first}
%\end{center}
%\end{figure*}

%-----------------IGR J14331-6112--------------------------------------------------------------------------------------

\subsection{IGR J14331-6112}\label{14331}

IGR J14331-6112 was discovered in 2006 by \citet{Keek2006} at the position (RA, Dec J2000.0) = (14$^h$33$^m$06$^s$.2, -61$^\circ$12\arcmin23\arcsec, 3\farcm9 uncertainty). \citet{Landi2007} obtained a \textit{Swift} position and detected a possible USNO-B1.0 counterpart, allowing \citet{Masetti2008} to carry out optical spectroscopy. From this, the authors suggested the source to be an HMXB hosting a companion star with a BIII or a BV spectral type. Finally, \textit{Chandra} observations conducted by \citet{Tomsick2009} confirmed the identification of the system and showed that the \textit{Chandra} energy spectrum is well fitted with an absorbed power law with a column density value N$_H$ = 2.2$^{+0.9}_{-0.8}\times$10$^{22}$ cm$^{-2}$ and a photon index $\Gamma$=0.34$_{-0.33}^{+0.37}$. Moreover, a very strong iron K$\alpha$-emission line is detected with an equivalent width of $\sim$ 945 eV. A possible existence of a soft excess is related. \\

The $K_s$ band spectrum of the source is shown in Figure \ref{second}. We report the detected lines in Table \ref{lines_all}. Even though the S/N of this spectrum is low, it shows the Br(7-4) emission line at 2.1657 $\mu$m. Moreover, assuming that the Br$\gamma$ emission line is mainly detected in supergiant and Be stars (hardly ever in main-sequence stars), and knowing that in supergiant star spectra, \ion{He}{i} at 2.058 $\mu$m in emission is generally detected as well, we argue that IGR J14331-6112 could be a Be star, since we do not detect this emission line. Thus, while we cannot conclude definitively on the spectral type of the companion star of IGR J14331-6112, it is probably a Be star, according to the presence of the Br(7-4) emission line.

%-----------------IGR J14488-5942--------------------------------------------------------------------------------------

\subsection{IGR J14488-5942}

IGR J14488-5942 was first mentioned by \citet{Bird2010} as a transient source. Two X-ray sources are detected in the \textit{\textit{\textit{INTEGRAL}}}/IBIS error circle by \citet{Landi2009} and during \textit{Swift} observations by  \citet{Rodriguez2010}. Both seem to be Galactic sources but the authors were unable to further confirm which of these two sources was the true counterpart of the \textit{\textit{\textit{INTEGRAL}}} detection. Both \citet{Landi2009} and \citet{Rodriguez2010} suggest that Swift J144843.3-594216 is the true \textit{\textit{\textit{INTEGRAL}}}/IBIS counterpart. 2MASS J14484322-5942137 seems to be the nIR counterpart with J=15.46, H=13.53 and $K_s$=12.43. Its \textit{Swift}/XRT spectrum is well fitted with an absorbed power law spectrum, with N$_H$ value of 1.3$\times$10$^{23}$ cm$^{-2}$, a photon index of $\sim$ 3 according to \citet{Landi2009} and \citet{Rodriguez2010}. Finally, \citet{Corbet2010} analyzed the 15-100 keV light curve of this source and detected a highly significant modulation at a period near 49 days. This long term variation is interpreted as the orbital period of an HMXB (possibly a BeHMXB). \\

The $K_s$-band spectrum of the source (see Figure \ref{second}) only shows potential emission lines of \ion{He}{i} at 2.0590 $\mu$m and Br(7-4) at 2.1665 $\mu$m. All the detected lines are reported in Table \ref{lines_all}. We cannot strictly conclude on the spectral type of the companion star but, as far as we know, there is no supergiant star that presents both \ion{He}{i} at 2.058 $\mu$m and Br$\gamma$ in emission with the same intensity and the absence of any other feature. On the contrary, \citealt{Clark1999} indicates that classical BeHMXB do not usually show evidence for emission from species other than \ion{H}{i} and \ion{He}{i}, and that the equivalent width of \ion{He}{i} at 2.058 $\mu$m typically exceeds that of Br$\gamma$. Since we only detected \ion{H}{i} and \ion{He}{i} and because the equivalent width of \ion{He}{i} at 2.058 $\mu$m is greater than that of Br$\gamma$, we conclude that this HMXB is more likely an Oe/Be HMXB than a supergiant one.

%-----------------IGR J16195-4945--------------------------------------------------------------------------------------

\subsection{IGR J16195-4945}\label{16195}

IGR J16195-4945 was detected with \textit{\textit{\textit{INTEGRAL}}} \citep{Walter2004}. \citet{Sidoli2005}, using \textit{\textit{\textit{INTEGRAL}}} observations, derive an average flux of $\sim$ 17 mCrab in the 20-40 keV band. Follow-up observations conducted by \citet{Sguera2006} showed that the source behaves like an SFXT and reported a peak-flux of $\sim$ 35 mCrab in the 20-40 keV band. \citet{Tomsick2006}, using \textit{Chandra} observations refined the position of the source to 0\farcs6 accuracy, which allowed them to find the nIR and mIR counterparts in the 2MASS (2MASS J16193220- 4944305) and in the GLIMPSE (G333.5571+00.3390) catalogs. The spectrum is well fitted with an absorbed power law with $\Gamma$ $\sim$ 0.5 and N$_{\rm{H}}$ $\sim$ 7 $\times$ 10$^{22}$ cm$^{-2}$. According to \citet{Tomsick2006}, the SED of the companion star is compatible with an O, B or A supergiant star. A possible USNO counterpart was detected but both \citet{Tomsick2006} and \citet{Tovmassian2006} suggested that the source is blended with a foreground object. \citet{Rahoui2008} did not detect the source in 2006 with VISIR, the mIR-infrared instrument on the VLT UT3 telescope. Nevertheless, they fitted its SED using nIR data and the GLIMPSE flux values given in \citet{Tomsick2006}. This study revealed that the source exhibits a mIR excess, which is needed to correctly fit the SED. However, this excess is weak and the stellar component is still consistent with an O/B massive star \citep{Rahoui2008}. Finally, \citet{Morris2009}, using \textit{Suzaku} observations, revealed a heavily absorbed HMXB and mentioned the possible presence of a disk around the donor star through models of accretion. From theoretical considerations, they can also estimated a tentative orbital period of about 16 days.\\

\begin{figure*}[h!]
\begin{minipage}[t]{1\textwidth}
\begin{center}
\includegraphics[width=0.5\textwidth]{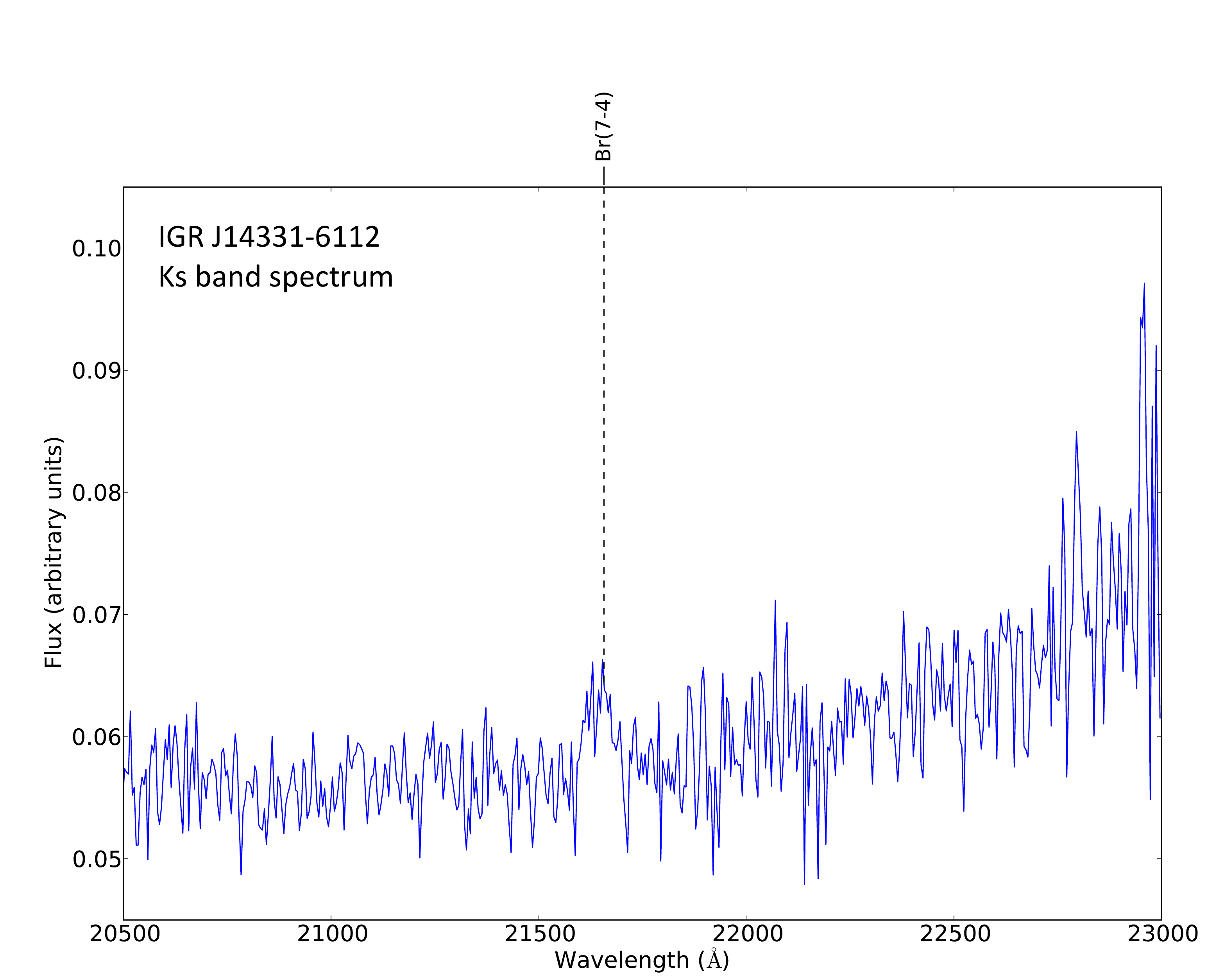}\hfill
\includegraphics[width=0.5\textwidth]{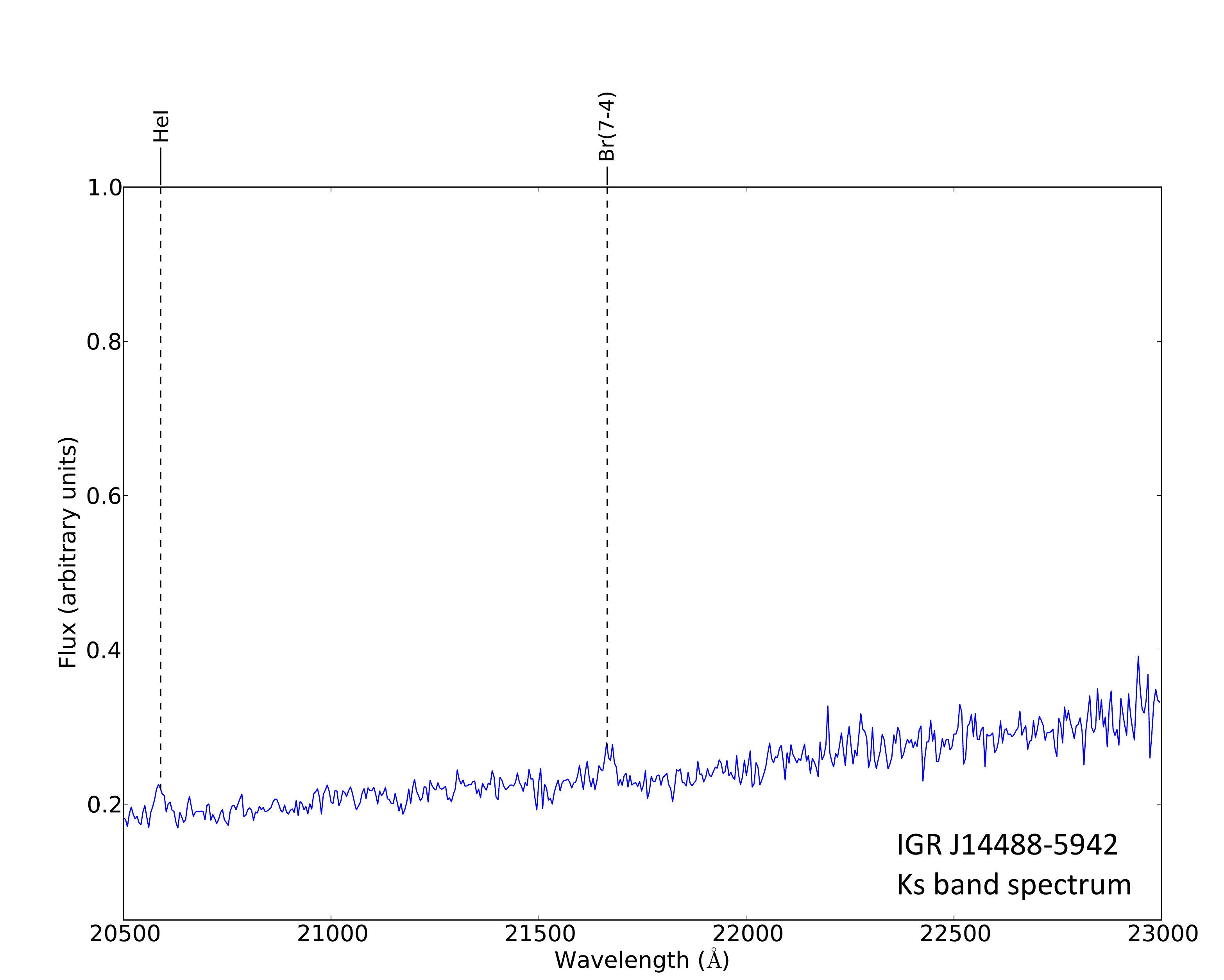}\\
\includegraphics[width=0.5\textwidth]{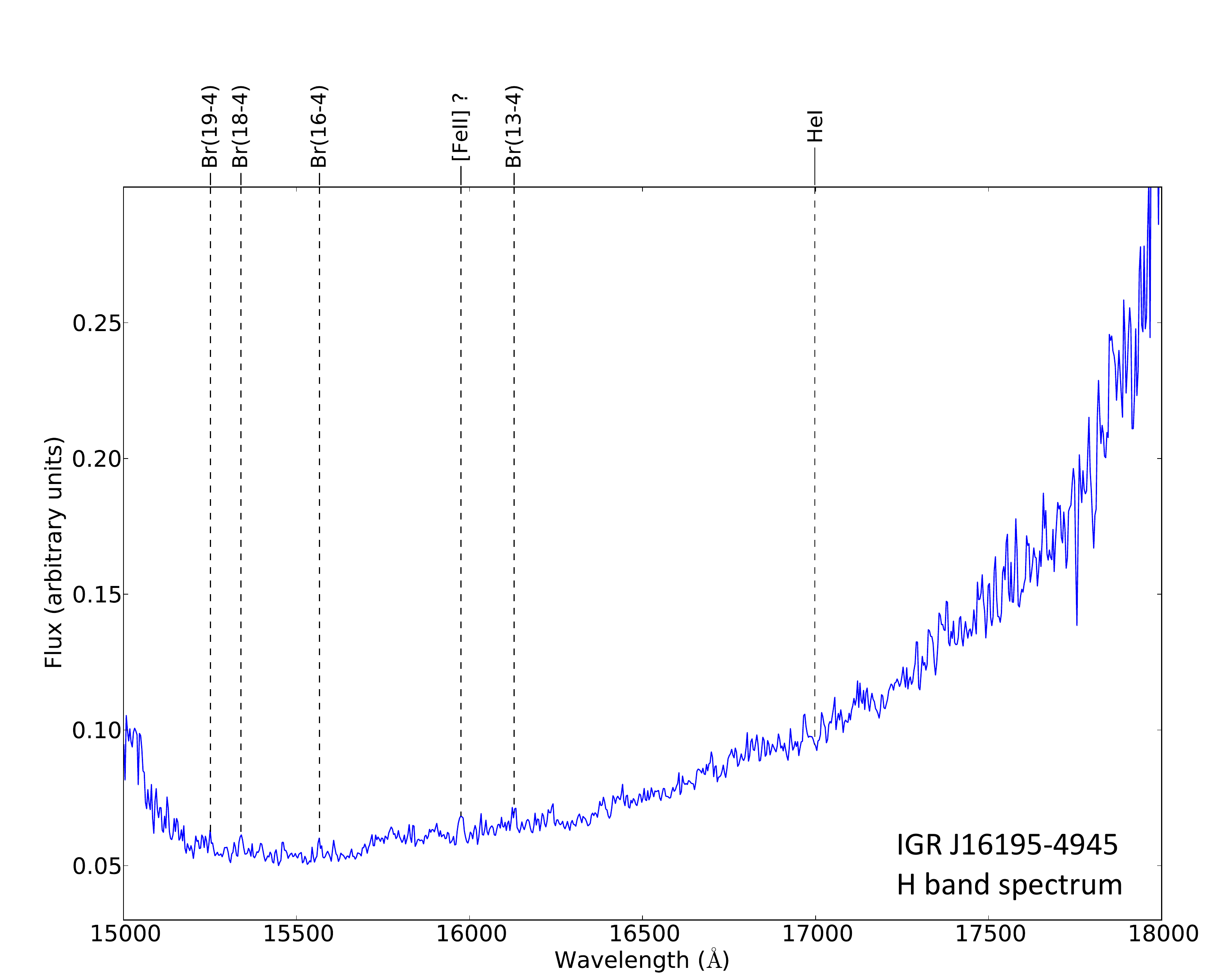}\hfill
\includegraphics[width=0.5\textwidth]{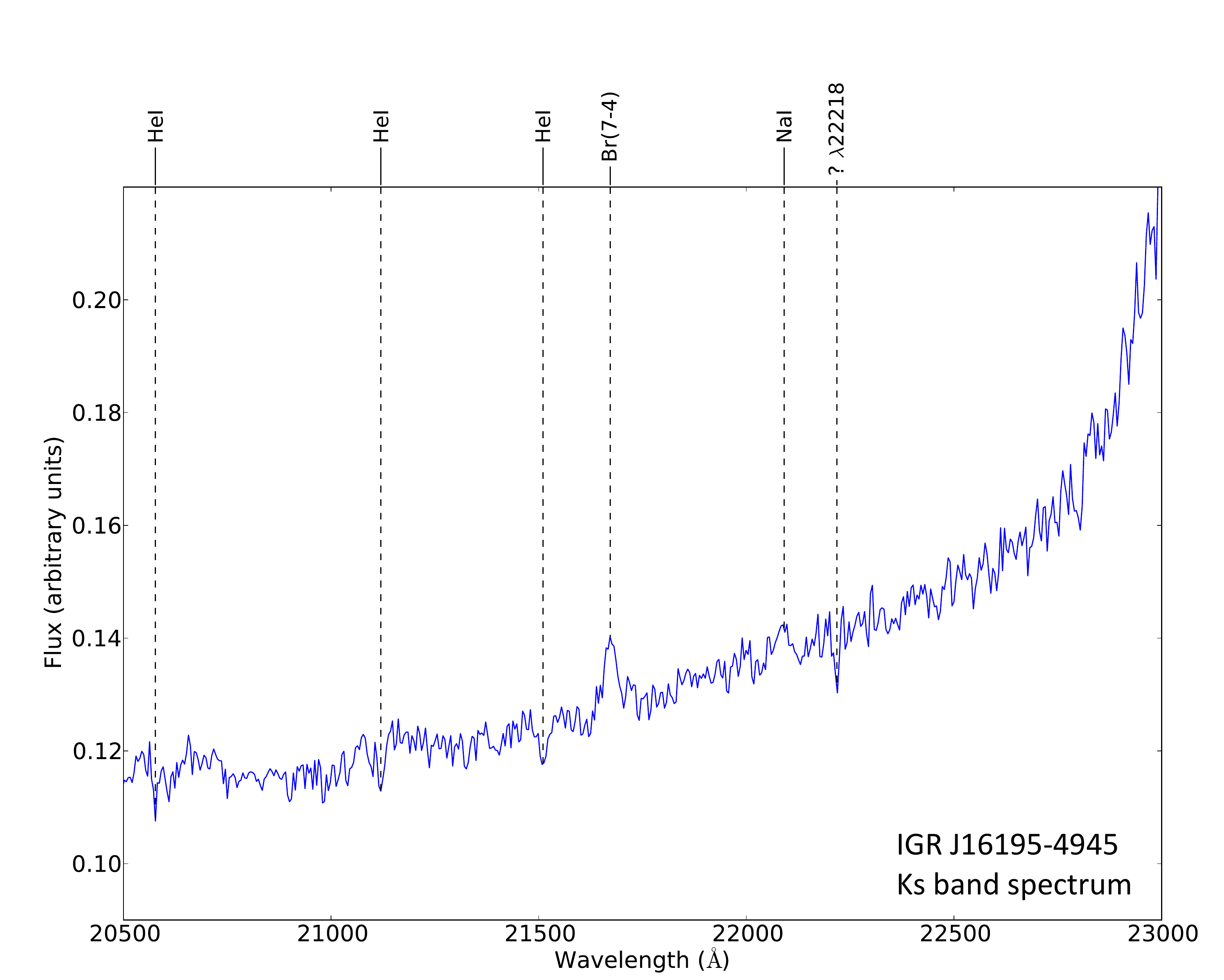} \\
\end{center}
\end{minipage}
\caption{Spectra of IGR J14331-6112 (top left), IGR J14488-5942 (top right), and IGR J16195-4945 (H band: bottom left and $K_s$ band: bottom right).} 
\label{second}
\end{figure*}

The $H$- and $K_s$-band spectra are shown in Figure \ref{second} and the detected lines are reported in Table \ref{lines_all}. The $H$-band spectrum is rather faint and exhibits very few lines. This is consistent with a supergiant late O-type star \citep[see e.g.][]{Morel2005, Hanson1998}. Moreover, the equivalent width of the \ion{He}{i} absorption at 1.700 $\mu$m is closer to the value expected for a supergiant star than for a dwarf star \citep{Hanson1998}. The $K_s$-band spectrum shows Br(7-4) emission and a 2.1120 $\mu$m absorption lines with large equivalent widths that are compatible with those observed in O9.7 supergiant stars \citep[see][]{Hanson1996}. Moreover, there are two other \ion{He}{i} absorption lines at 2.0577 $\mu$m and 2.150 $\mu$m which lead to an ON9.7Iab stellar classification \citep{Hanson1996}.

%\begin{figure*}[th!]
%\begin{center}
%\includegraphics[scale=0.25]{secondgroup.pdf}
%\caption{Spectra of IGR J14331-6112, IGR J14488-5942, IGR J16195-4945. }
%\label{second}
%\end{center}
%\end{figure*}

%-----------------IGR J16318-4848--------------------------------------------------------------------------------------

\subsection{IGR J16318-4848}\label{16318}

IGR J16318-4848 is the first source discovered with \textit{\textit{\textit{INTEGRAL}}} by the ISGRI detector on 20 January 2003 \citep{Courvoisier2003}. ToO observations, triggered with \textit{XMM-Newton} provided a more accurate localization \citep{Walter2003} and showed that the source was exhibiting a strong column density N$_H \sim 2\times10^{24}$ cm$^{-2}$ \citep{Matt2003,Walter2003}. Moreover, highly variable flux (by a factor of 20) with a time lapse of ten hours followed by two to three days of inactivity are signatures of wind accretion, reminiscent of other peculiar high energy sources such as XTE J0421+560/CI Cam and GX 301-2 \citep{Revnivtsev2003}. Using the \textit{XMM-Newton} position, \citet{Filliatre2004} triggered ToO optical and nIR observations just after the detection of the source that led to the discovery of the counterpart, extremely bright in nIR and highly absorbed in the optical V-band (A$_V$=17.4 mag). Nevertheless, this absorption in visible is 100 times lower than the absorption in X-rays. Thus, \citet{Filliatre2004} suggested that the material absorbing in the X-rays had to be concentrated around the compact object, while the material absorbing in optical/nIR would extend around the whole binary system. 
The nIR spectrum is very rich in many strong emission lines, originating from different media, that suggests the presence of a stratified circumstellar environment. This allowed \citet{Filliatre2004} to suggest the companion star to be a sgB[e], which are stars with extreme environments. Its position in the Hertzsprung-Russel diagram, at the edge of the blue supergiant domain, also confirms the extreme behavior of this object. Archival mid-infrared (mIR) data from \textit{Spitzer} GLIMPSE survey showed a long wavelength ($\lambda \geqslant$ 4 $\mu$m) excess. Fitting this excess with a blackbody model, \citet{Kaplan2006} concluded that it might be due to warm circumstellar dust emission. Subsequent \textit{Spitzer}/IRS mIR spectroscopic observations, conducted by \citet{Moon2007} confirmed the hot (T $>$ 700K) circumstellar dust and also  suggested a warm (T$\sim$190K) dust component. \citet{Rahoui2008}, using mIR photometric observations with VLT/VISIR, built the optical to mIR SED and suggested a cocoon of dust that is enshrouding the whole binary system. With new VISIR mIR spectroscopic observations and using archival NTT + Spitzer observations,\citet{Chaty2012} recently excluded the spherical geometry for the dust component and the warm dust component suggested by \citet{Moon2007} by fitting broadband nIR to mIR SED. Indeed, using an emission model of intermediate mass star such as Herbig Ae/Be stars adapted for sgB[e] stars, they showed the presence of two-temperature components in a toric geometry, with an irradiated toric rim of temperature $T_{\rm{rim}}$ $\sim$ 3800--5500 K, surrounded by a hot dusty viscous disk component at $T_{\rm{dust}}$ $\sim$ 767 -- 923 K with a maximum extension of $r_{out}$ = 5.6 au.\\

We present our $K_s$-band spectrum in Figure \ref{three} and detected lines are reported in Table \ref{lines_all_2}. It exhibits many intense and broad emission lines that were identified mainly according to \citet{Morris1996} and \citet{McGregor1988}. Two lines (\ion{He}{i} at 2.058 $\mu$m and \ion{He}{i} at 2.1126 $\mu$m) show a P-Cygni profile, confirming that there is a strong stellar wind of circumstellar material. From these lines, we derived a wind speed value of $\sim$ 400 km s$^{-1}$ consistent with the value previously derived by \citet{Filliatre2004}. As underlined in \citet{Filliatre2004}, many lines are common to CI Cam \citep{Clark1999} indicating similar physical conditions. In contrast to the CI Cam and to the IGR J16318-4848 nIR spectra published by \citet{Filliatre2004}, we detect in our $K_s$-band spectrum a line that could be an H$_2$ transition at 2.1218 $\mu$m, suggesting that either shock heating is high enough in this source or that there is a region of sufficiently low temperature where molecules such as H$_2$ are not dissociated \citep[see][]{Clark1999}. Table \ref{16318_var} compares the equivalent width (EW) and FWHM values between data of 2003 \citep{Filliatre2004} and data presented in this study, obtained in 2010. Assuming 20$\%$ of uncertainty for our EW and FWHM measurements (as noted above, and estimated after repeting the measurement process with the IRAF \textit{splot} task), no major variation is detected for the brightest lines. Other lines such as $\left[\rm{\ion{Fe}{ii}}\right]$ at 2.046 $\mu$m, \ion{He}{i} at 2.1126 $\mu$m, \ion{N}{iii} + \ion{C}{iii} at 2.116 $\mu$m, \ion{He}{i} at 2.1847 $\mu$m, and $\left[\rm{\ion{Fe}{ii}}\right]$ at 2.224 $\mu$m, exhibit variations between 25 and 65$\%$, but their faintness can engender more important uncertainties on EW and FWHM values, which prevents any further conclusion on these line variabilities.These observations confirm the sgB[e] nature of IGR J16318-4848.

%-----------------IGR J16320-4751--------------------------------------------------------------------------------------

\subsection{IGR J16320-4751}

IGR J16320-4751 was discovered in February 2003 by \citet{Tomsick2003} at the position RA = 16$^{\rm{h}}$32$^{\rm{m}}$.0, Dec = -47$^\circ$51\arcmin (equinox J2000.0; uncertainty 2\farcm0). Follow-up \textit{XMM-Newton} observations localized the source at RA = 16$^{\rm{h}}$32$^{\rm{m}}$01$^{\rm{s}}$.9, Dec = -47$^\circ$52\arcmin27\arcsec (equinox J2000.0; uncertainty 3\arcsec) \citep{Rodriguez2003, Rodriguez2006}. This result agrees with \citet{Negueruela2007}. It is a heavily absorbed variable source with N$_{\rm{H}}$ $\sim$ 2.1 $\times$ 10$^{23}$ cm$^{-2}$ and a hard X-ray spectrum fitted by an absorbed power-law, with $\Gamma$ $\sim$ 1.6 \citep{Rodriguez2006}. Soft X-ray pulsations have been detected from this source at a period of P $\sim$ 1309$\pm$40 s with \textit{XMM-Newton} and P $\sim$ 1295$\pm$50\,s with \textit{ASCA}; these pulsations are the signature of an X-ray pulsar \citep{Lutovinov2005}. An orbital period of 8.96 $\pm$ 0.01 days was found from a \textit{Swift}/BAT lightcurve extending from 2004 December 21 to 2005 September 17 \citep{Corbet2005}, and of 8.99 $\pm$ 0.05 days with \textit{\textit{\textit{INTEGRAL}}} \citep{Walter2006}. Plotting the spin and orbital periods of this source on a Corbet diagram \citet{Corbet2005_1} suggests a supergiant HMXB nature. IGR J16320-4751 might have been persistent for at least eight years, since this source is the rediscovery of a previously known \textit{ASCA} source AX J1631.9-4752. NIR observations of the most likely counterpart, conducted by \citet{Chaty2008}, showed that the blue nIR spectrum presents only an Fe line, probably because it is very faint and absorbed. The red nIR spectrum exhibits a very red continuum and the presence of absorption and emission lines: the Pa(7-3) emission line, the Brackett series with P-Cygni profiles between 1.5 and 2.17 $\mu$m and He I at 2.166 $\mu$m (perhaps with P-Cygni profile). According to \citet{Chaty2008}, these narrow and deep Paschen and He I lines, associated with P-Cygni profiles, are typical of early-type stars, and more precisely of luminous supergiant OB stars, which is therefore the most likely spectral type of the companion star. The wide Br$\gamma$ line would constrain the spectral type to an O supergiant or even O hypergiant. SED fitting computed in \citet{Rahoui2008} and observations of \citet{Chaty2008} finally state that this source belongs to the very obscured supergiant HMXB class, hosting a neutron star. \\

In our $K_s$-band spectrum (see Figure \ref{three} and the detected lines in Table \ref{lines_all_2}), the emission line of Br(7-4) is clearly detected at 2.1668 $\mu$m with possibly a P-Cygni profile. The \ion{He}{i} emission line at 2.0586 $\mu$m is weak but quite broad, whereas another \ion{He}{i} absorption line is detected at 2.1127 $\mu$m. These lines lead to a classification of the companion star as a BN0.5Ia according to \cite{Hanson2005}, confirming the supergiant type of this HMXB. We point out that the spectrum exhibits numerous absorption lines that we were unable to associate with physical transitions, although we checked that these features do not come from a poor telluric correction.

%-----------------IGR J16328-4726--------------------------------------------------------------------------------------

\subsection{IGR J16328-4726}

IGR J16328-4726 was discovered with \textit{\textit{\textit{INTEGRAL}}} by \citet{Bird2007}. A flare, observed by \citet{Grupe2009}, was detected with \textit{Swift}/XRT on 2009 June 10 at the coordinates RA = 16$^h$32$^m$37$^s$.88, Dec= -47$^\circ$23\arcmin42\farcs4,  (equinox J2000.0; uncertainty 1\farcs7). The XRT light curve shows that the source brightness is steadily decaying. Moreover, the X-ray spectrum in the XRT waveband can be fitted by an absorbed power law model with $\Gamma$=0.56$^{+0.75}_{-0.68}$ and an absorption column density N$_H$ = 8.1$^{+5.7}_{-4.9}\times10^{22}$ cm$^{-2}$ which tends to show an absorption excess in comparison with the expected galactic value. Because of this high absorption, \citet{Grupe2009} were unable to identify an optical counterpart. However, a 2MASS counterpart candidate is suggested at RA = 16$^h$32$^m$37$^s$.91 Dec = -47$^\circ$23\arcmin40\farcs9 (equinox J2000) with J=14.631, H=12.423, K=11.275 mag. \citet{Corbet2010_1} reported the analysis of the \textit{Swift}/BAT 58-month survey light curve of the source in the energy range 15 - 100 keV that reveals highly significant modulation at a period near 10-days. The mean flux is approximatively of 1.3 mCrab. The 10 day period suggests that the source is an HMXB, in particular one powered by accretion from the wind of a supergiant companion star. According to \citet{Corbet2010_1}, this classification would be consistent with the high level of absorption found by \citet{Grupe2009}. \citet{Fiocchi2010} used the \textit{\textit{\textit{INTEGRAL}}}/IBIS-JEM-X public database and XRT observations performed during a flare of the source. Both spectral and timing properties observed during outburst, its Galactic plane location and the presence of an IR star in the \textit{Swift}/XRT error circle suggest that this source is an SFXT. According to \citet{Fiocchi2010}, the source 2MASS J16323791-4723409 is the most likely candidate IR counterpart. \textit{XMM-Newton} observations, conducted by \citet{Bozzo2012}, revealed a variability on timescales of hundreds of seconds, typical of SFXT prototypes. Finally, \citet{Fiocchi2013} identified the counterpart as a high-mass OB type star, classifying this source as a firm HMXB.\\

Our $K_s$-band spectrum (see Figure \ref{three} and Table \ref{lines_all_2}) shows that the Br(7-4) emission line at 2.1661 $\mu$m is clearly detected. Moreover, \ion{He}{i} at 2.0579 $\mu$m and at 2.1114 $\mu$m are detected in absorption, together with \ion{N}{iii}/\ion{C}{iii} emission around 2.116 $\mu$m. According to \cite{Hanson1996}, it can be an O8Iaf or more probably an O8Iafpe, taking into account the \ion{He}{i} / Br(7-4) line ratio and the \ion{N}{iii}/\ion{C}{iii} emission, which are typically observed in supergiant star spectra. Moreover, the \ion{C}{iv} line also strengthens the supergiant classification of this system. This classification agrees with previous publications which suggested this source is an sgHMXB/SFXT.

%-----------------IGR J16418-4532--------------------------------------------------------------------------------------

\subsection{IGR J16418-4532}

IGR J16418-4532 was discovered in 2004 by \citet{Tomsick2004} with \textit{\textit{\textit{INTEGRAL}}} at the position RA = 16$^{\rm{h}}$41$^{\rm{m}}$.8, Dec = -45$^\circ$32\arcmin (equinox J2000.0; uncertainty 2\farcm0). \textit{XMM-Newton} localized the source at RA = 16$^{\rm{h}}$41$^{\rm{m}}$51$^{\rm{s}}$.0, Dec = -45$^\circ$32\arcmin25\arcsec (equinox J2000.0; uncertainty 4\arcsec) \citep{Walter2006}. These observations have shown that it is a heavily absorbed X-ray pulsar exhibiting a column density of N$_H \sim 1.0\times10^{23}$ cm$^{-2}$, a peak-flux of $\sim$ 80 mCrab in the 20--30 keV band, and a pulse period of 1246 $\pm$ 100 s \citep{Walter2006}. \citet{Sguera2006}, using \textit{\textit{\textit{INTEGRAL}}} observations, suggested that this source is an SFXT candidate. A 3.75-day modulation was found in the \textit{Rossi-XTE}/ASM and \textit{Swift}/BAT lightcurves, with a possible total eclipse, which would suggest either a high binary inclination, or a supergiant companion star \citep{Corbet2006}. The latter case would be consistent with the position of the object in the Corbet diagram \citep{Chaty2008}. NIR observations conducted by \citet{Chaty2008} proposed four nIR  candidate counterparts and the brightest one, 2MASS J16415078-4532253, is favored to be the candidate counterpart. SED fitting carried out by \citet{Chaty2008} suggested an OB spectral type companion star and the source would be a supergiant HMXB. But its X-ray behavior may rule out that  this source belongs to the SFXT class. \citet{Rahoui2008} carried out SED fitting using nIR and mIR observations which gave an absorption in the V-band value of A$_V$=14.5 even though the enshrouding material marginally contributes to its mIR emission. In February 2011, \citet{Romano2012} performed \textit{Swift}/XRT observations confirming that the spectrum is well fitted by an absorbed power law. The source is quite bright, with maybe the signature of a flare characteristic of SFXT \citep[see e.g.][]{Romano2013}. The 40 ks \textit{XMM-Newton} observations carried out in February 2011 by \citet{Sidoli2012} show strong variability of two orders of magnitude with several bright flares. The type of X-ray variability displayed by IGR J16418-4532, its dynamic range and timescale, and the quasi-periodic flaring are all suggestive of a transitional accretion regime between pure wind accretion and full Roche lobe overflow according to \citet{Sidoli2012}. Finally, \citet{Drave2013} used combined \textit{INTEGRAL} and \textit{XMM-Newton} observations to reveal an X-ray intensity dip in this pulsating SFXT, which may be explained by a highly magnetized neutron star.\\

\begin{figure*}[h!]
\begin{minipage}[t]{1\textwidth}
\begin{center}
\includegraphics[width=0.5\textwidth]{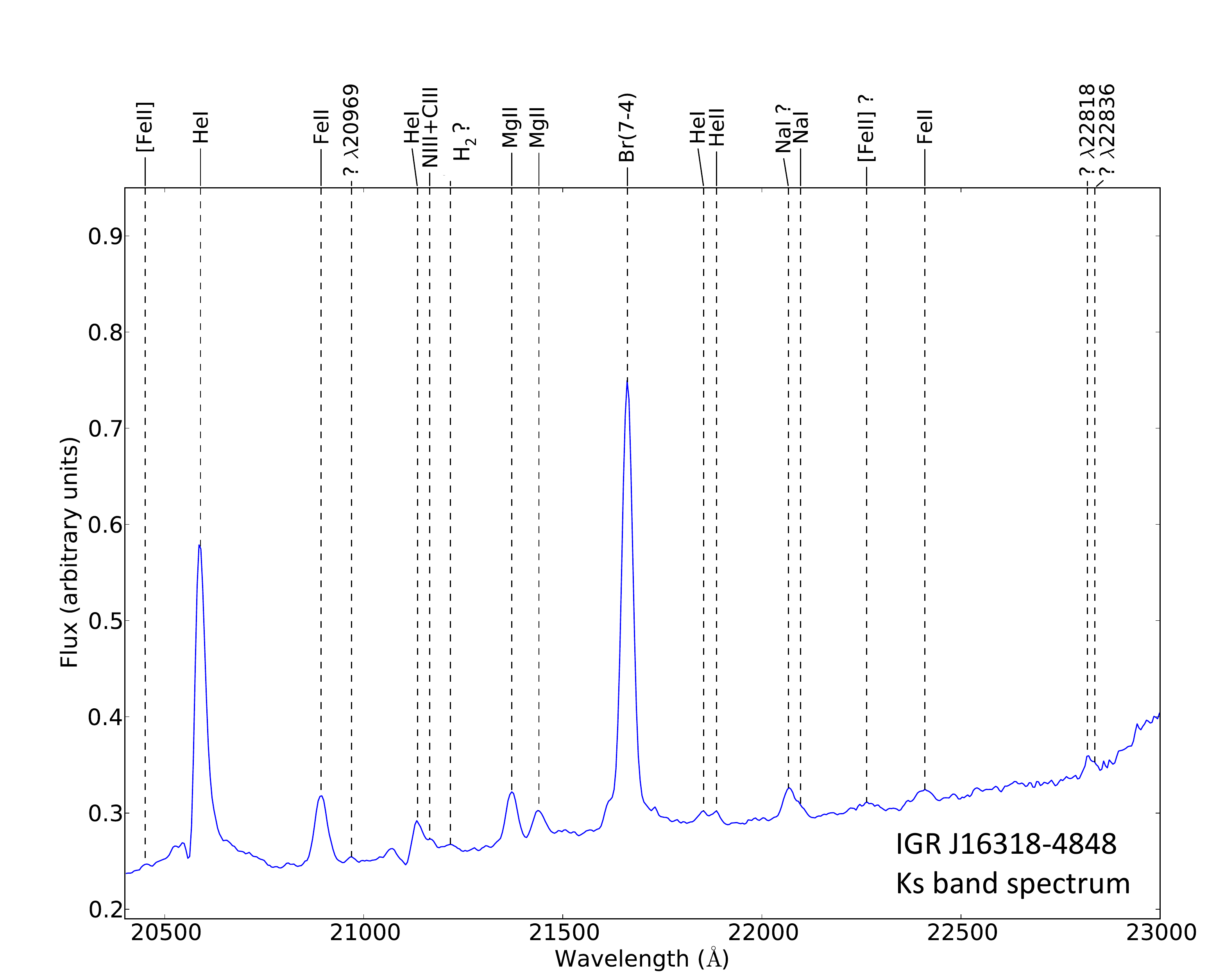}\hfill
\includegraphics[width=0.5\textwidth]{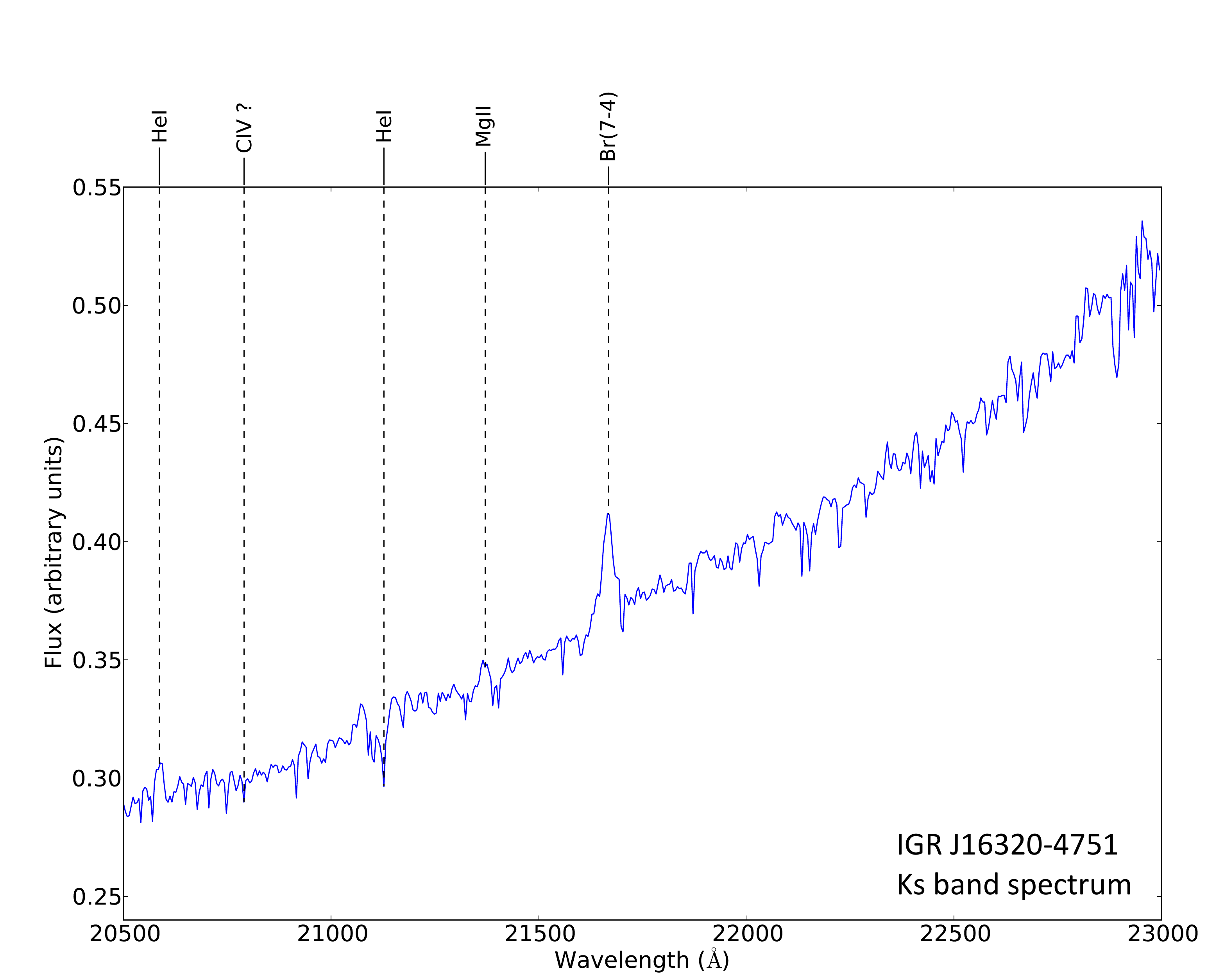}\\
\includegraphics[width=0.5\textwidth]{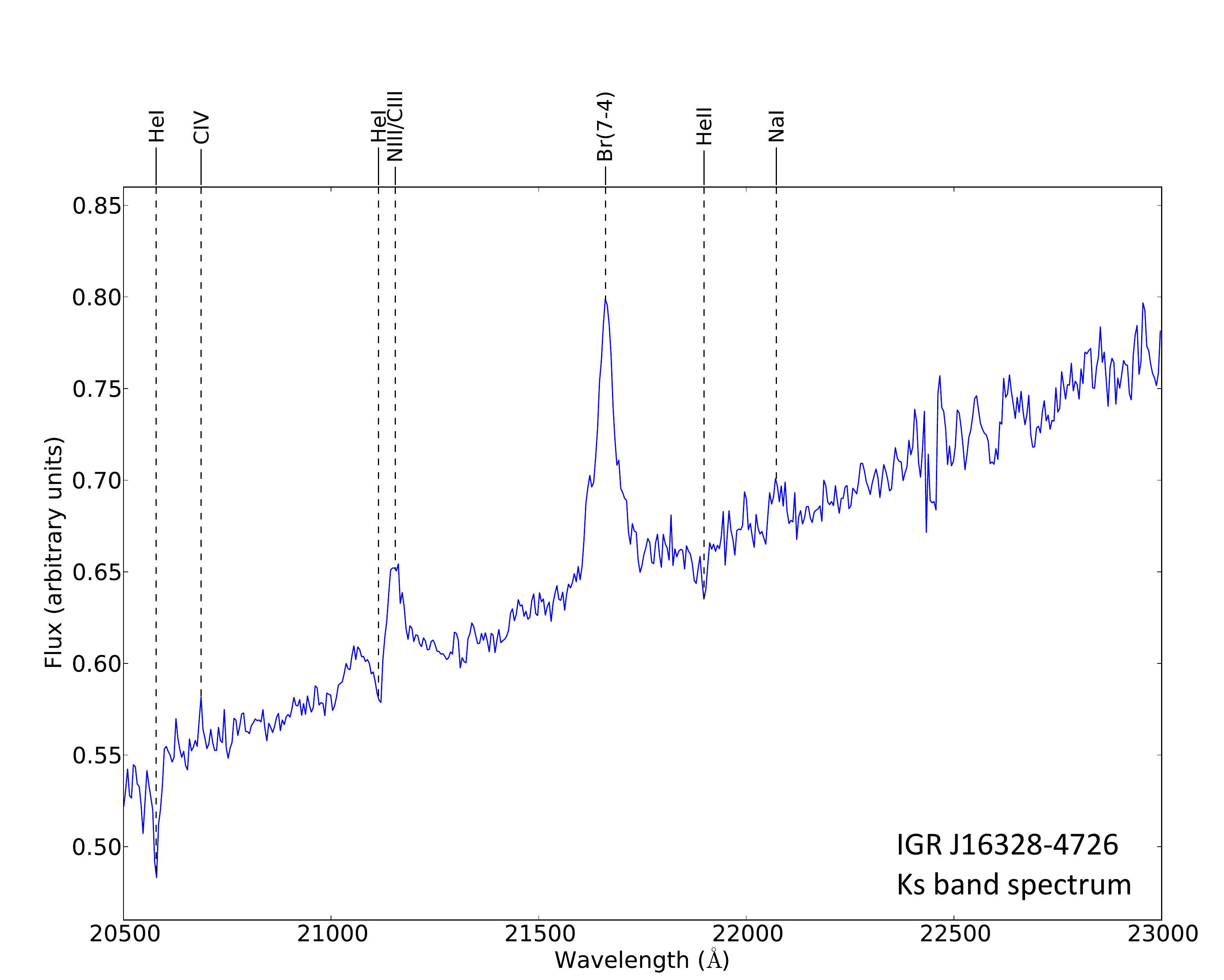}\hfill
\includegraphics[width=0.5\textwidth]{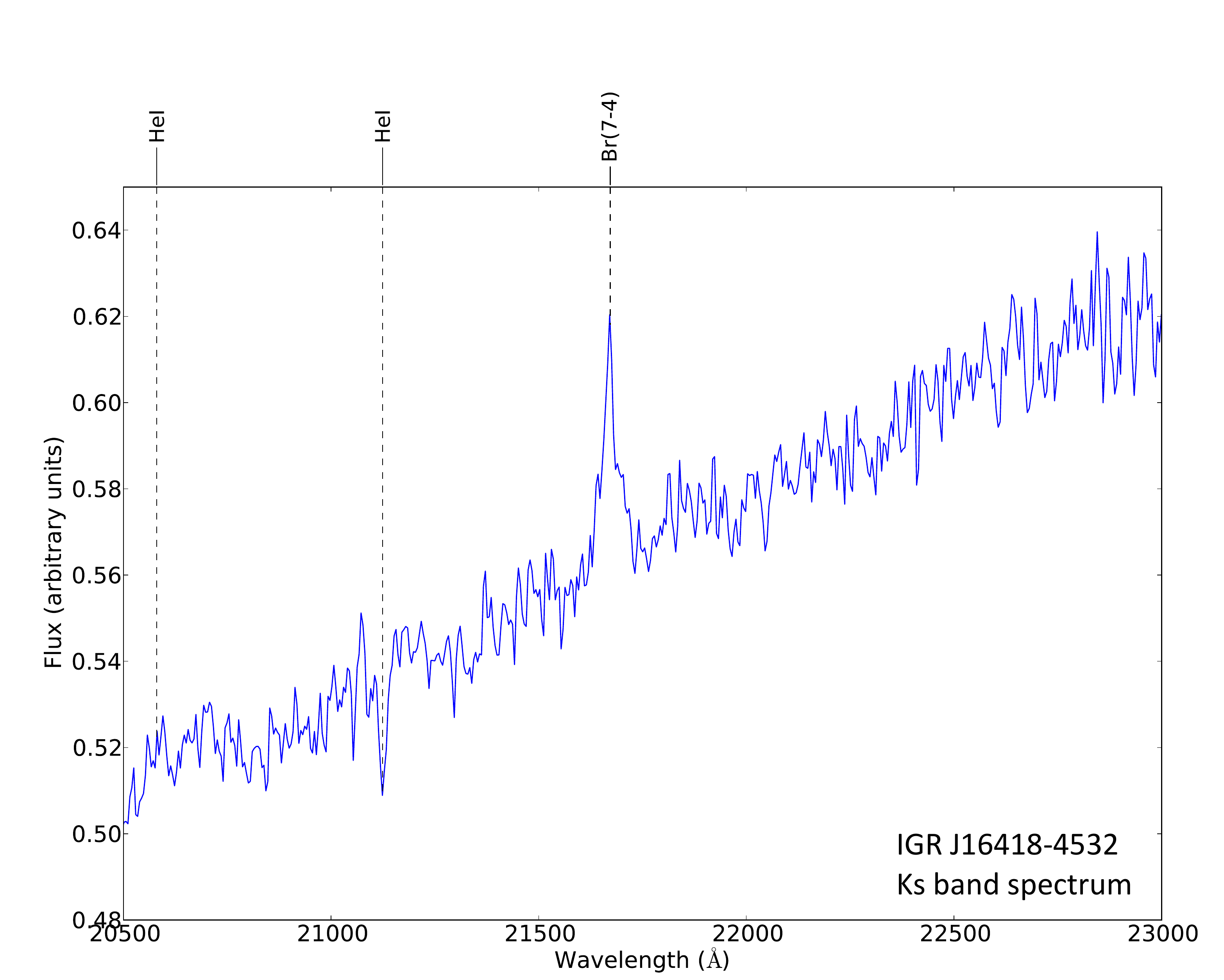} \\
\end{center}
\end{minipage}
\caption{Spectra of IGR J16318-4848 (top left), IGR J16320-4751 (top right), IGR J16328-4726 (bottom left), and IGR J16418-4532 (bottom right).} 
\label{three}
\end{figure*}

\begin{figure*}[th!]
\begin{center}
\includegraphics[scale=0.5]{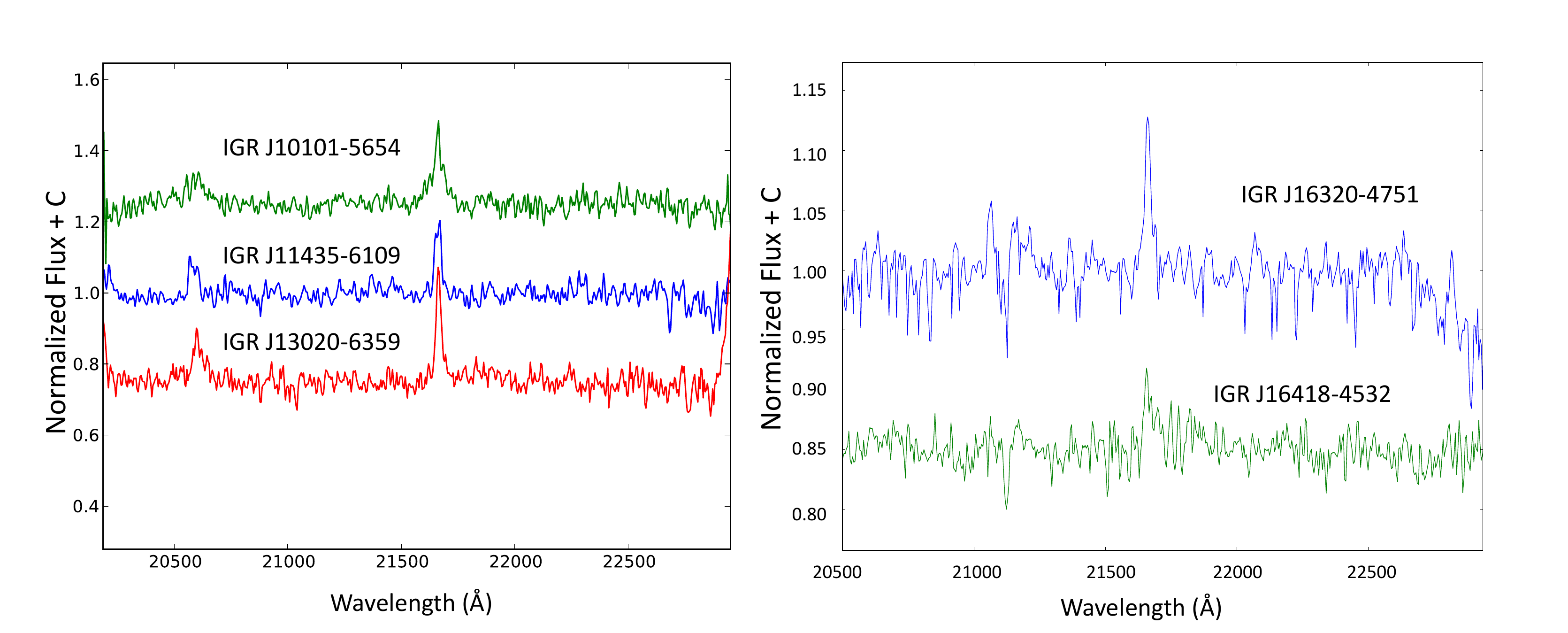}
\caption{Left panel: comparison of the $K_s$-band spectra of IGR J10101-5654, IGR J11435-6109, and IGR J13020-6359 (spectra are normalized and offset). Right panel: comparison of the $K_s$-band spectra of IGR J16320-4751, and IGR J16418-4532 (spectra are normalized and offset).}
\label{comparison}
\end{center}
\end{figure*}

The nIR $K_s$-band spectrum is presented in Figure \ref{three} and Table \ref{lines_all_2}. It shows a wide emission line at 2.1672 $\mu$m that corresponds to the Br(7-4) line and a weak emission line of \ion{He}{i} at 2.0580 $\mu$m. Moreover, we observed an absorption line of \ion{He}{i} at 2.1124 $\mu$m. These detections point toward a BN0.5Ia spectral type\footnote{We point out that X-shooter observations \citep[see][]{Goldoni2013} show Br(7-4) line in absorption and then lead to an O9.5I spectral classification. Nevertheless, this result confirms the supergiant classification that we propose for this HMXB.} (see Figure \ref{comparison}, right panel, for a comparison with the spectrum of IGR J16320-4751).

%
%\begin{figure*}[th!]
%\begin{center}
%\includegraphics[scale=0.25]{thirdgroup.pdf}
%\caption{Spectra of IGR J16318-4848, IGR J16320-4751, IGR J16328-4726 and IGR J16418-4532. }
%\label{three}
%\end{center}
%\end{figure*}

%instrument, date + equinoxe + check position ???
%Zurita Heras & Chaty in prep ? (X-ray behaviour cf Chaty 2008)
%Citation Corbet2006 !!

%\subsubsection*{K band}

%$\sigma_{noise}=0.0119$.\\
%Only the emission line of Br(7-4) at 2.1661 $\mu$m is detected at 5$\sigma$. The absorption line of \ion{He}{i} at 2.1126 $\mu$m is detected above 3$\sigma$. 
%Seems to be the spectrum of a \textbf{BN0.5Ia} star.

%no confidence in the detection of \ion{He}{i} at 2.058 $\mu$m.
%gaussian fit complicated for Br(7-4)
%line around 2.18 $\mu$m visible in raw data...

%-----------------IGR J17200-3116---------------------------------------------------------------------------------------

\subsection{IGR J17200-3116}

IGR J17200-3116 was discovered with \textit{\textit{\textit{INTEGRAL}}} by \citet{Walter2004}. \textit{ROSAT} observations carried out by \citet{Stephen2005} refined the source coordinates and localized a possible USNO B1 counterpart with B $\sim$ 19. The optical spectrum obtained by \citet{Masetti2006} is typical of the HMXB class, clearly showing a narrow H$\alpha$ line and a strongly reddened continuum. Moreover, \citet{Tomsick2008}, using \textit{Chandra} observations confirm the counterpart candidate suggested by \citet{Masetti2006}. The \textit{Chandra} spectrum was fitted by an absorbed power law with $\Gamma$=0.8 and N$_H = 1.9\times10^{23}$ cm$^{-2}$. This column density does not require local absorption. Finally, \citet{Nichelli2011}, thanks to \textit{Swift}-XRT X-ray light curves, detected coherent pulsations at a period of 328.182(3) s. This periodicity is confirmed in the 18-30 keV range by \textit{INTEGRAL/ISGRI} archival observations from 2003 to 2006. All these results confirm the HMXB type of IGR J17200-3116.\\

The lack of optical photometry of this source prevents us from fitting the SED. In addition, due to bad weather, the NIR spectra are not useful. Thus, more observations are required to determine the spectral type of this HMXB.

%-----------------IGR J17354-3255--------------------------------------------------------------------------------------

\subsection{IGR J17354-3255}
%Kulkeers 2006 dŽcouverte ??
IGR J17354-3255 was discovered with \textit{\textit{\textit{INTEGRAL}}} by \citet{Bird2007}. The \textit{Swift} and \textit{Chandra} observations carried out by \citet{Vercellone2009} and \citet{Tomsick2009} allowed the authors to locate the counterpart (2MASS J17352760-3255544) and show that the source is well fitted by an absorbed power law in the 0.3-10 keV range. \citet{Tomsick2009} suggest that the source is an HMXB. Moreover, \citet{Dai2011}, using the \textit{Swift/BAT} instrument, unveiled an 8.448 $\pm$ 0.002 days periodicity that is identified with the orbital period of the system, typical of a wind accretor X-ray binary. They also confirmed that the X-ray spectrum of the source is compatible with an HMXB X-ray spectral emission. Carrying out a long-term \textit{\textit{\textit{INTEGRAL}}} monitoring of the source and using \textit{Swift} archival observations, \citet{Sguera2011} showed that IGR J17354-3255 is a weak and persistent source that emits occasional X-ray flares. They also confirmed the orbital period of 8.4 days and claim that the spectral and temporal characteristics of the source are indicative of an sgHMXB, whereas the dynamic ranges at soft and hard X-rays tend to be typical of intermediate SFXT. Finally, \citet{Bozzo2012}, using \textit{XMM Newton} spectra, claimed the source to be an SFXT.\\

The nIR $K_s$-band spectrum, presented in Figure \ref{fourth} and Table \ref{lines_all_2}, shows a clear \ion{He}{i} absorption at 2.1134 $\mu$m. A weak absorption is detected around 2.06 $\mu$m that could be an \ion{He}{i} absorption. Thus, this spectrum points toward an O8.5Iab(f) \citep[see][]{Hanson1996} or an O9Iab \citep[see][]{Hanson2005} spectral type and confirms the sgHMXB/SFXT classification of the binary source.

%-----------------IGR J17404-3655--------------------------------------------------------------------------------------

\subsection{IGR J17404-3655}

IGR J17404-3655 was first mentioned by \citet{Bird2007}. \citet{Landi2008_1}, with a series of X-ray follow-up observations performed with \textit{Swift/XRT}, determined an USNO and 2MASS counterpart. Moreover, \citet{Masetti2008}, by carrying out spectroscopic observations, showed that the counterpart (USNO-A2.0.0525\_28851523) presents a red continuum and a single narrow emission line consistent with the H$\alpha$ line with an equivalent width of $\sim$17 \AA. They concluded that the source is a Galactic X-ray source, most likely a low-mass X-ray binary at a distance of 9.1 kpc \citep{Masetti2009}. \citet{Masetti2009} ruled out an HMXB nature for this source because the optical magnitudes do not fit any early spectral type. However, \textit{Chandra} observations led by \citet{Tomsick2009} show that the spectrum is well fitted by an absorbed power law with a power-law index $\Gamma \sim -0.30^{+0.30}_{-0.24}$ which would be very unusual for an LMXB. Instead, these authors suggested that this source is an HMXB.\\

Only the Br(7-4) line is detected in emission in our nIR $K_s$-band spectrum (see Figure \ref{fourth} and Table \ref{lines_all_2}). %According to the high equivalent width value of the Br$\gamma$ emission line, it could be a supergiant star. 
As for IGR J14331-6112 (see Section \ref{14331}), assuming that the Br$\gamma$ line is mainly detected in supergiant and Be stars (hardly ever in main-sequence stars), and knowing that in supergiant star spectra, \ion{He}{i} at 2.058 $\mu$m in emission is generally detected as well, we argue that IGR J17404-3655 could be a Be star, since we do not detect any \ion{He}{i} emission line at 2.058 $\mu$m. The low S/N prevents any firm classification, but if the source is an HMXB, it would probably be a BeHMXB. However, we cannot rule out the LMXB classification yet.

%-----------------IGR J17586-2129--------------------------------------------------------------------------------------

\subsection{IGR J17586-2129}\label{17586}
IGR J17586-2129 was discovered by \citet{Bird2007}. \citet{Tomsick2009} localized the infrared 2MASS counterpart thanks to \textit{Chandra} observations. The X-ray spectrum is well fitted by a power-law with N$_H$ = 15.6 $\times$ 10$^{22}$ cm$^{-2}$ and $\Gamma \sim 0.23$. The authors stated that the source is a candidate absorbed HMXB. \textit{Swift} observations conducted by \citet{Krimm2009} during an outburst confirmed the association proposed by \citet{Tomsick2009}. They fitted the data with a power-law with N$_H$ = 1.11 $\times$ $10^{23}$ cm$^{-2}$ and $\Gamma \sim 1.14$. Finally, \citet{Sanchez-Fernandez2009}, using \textit{\textit{\textit{INTEGRAL}}} and \textit{Swift/BAT} observations, derived an average flux over seven days of 11.2 mCrab in the 18-40 keV band. The 18-150 keV spectra were well fitted by a power-law with $\Gamma=3.0$. Finally, if the spectrum remains unchanged, \citet{Sanchez-Fernandez2009} suggest that there is a cut-off in the 10-20 keV spectrum range.\\

\begin{figure*}[h!]
\begin{minipage}[t]{1\textwidth}
\begin{center}
\includegraphics[width=0.5\textwidth]{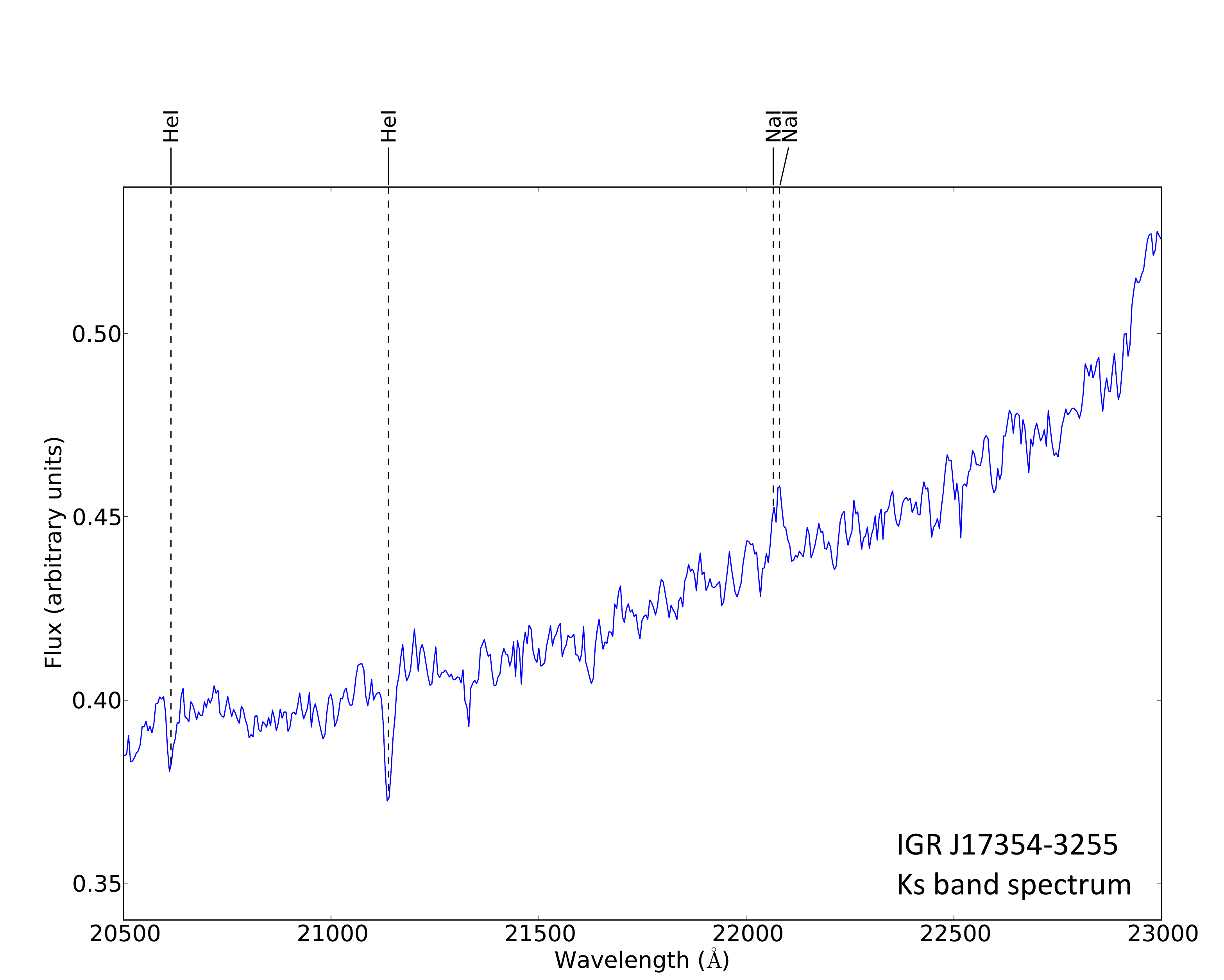}\hfill
\includegraphics[width=0.5\textwidth]{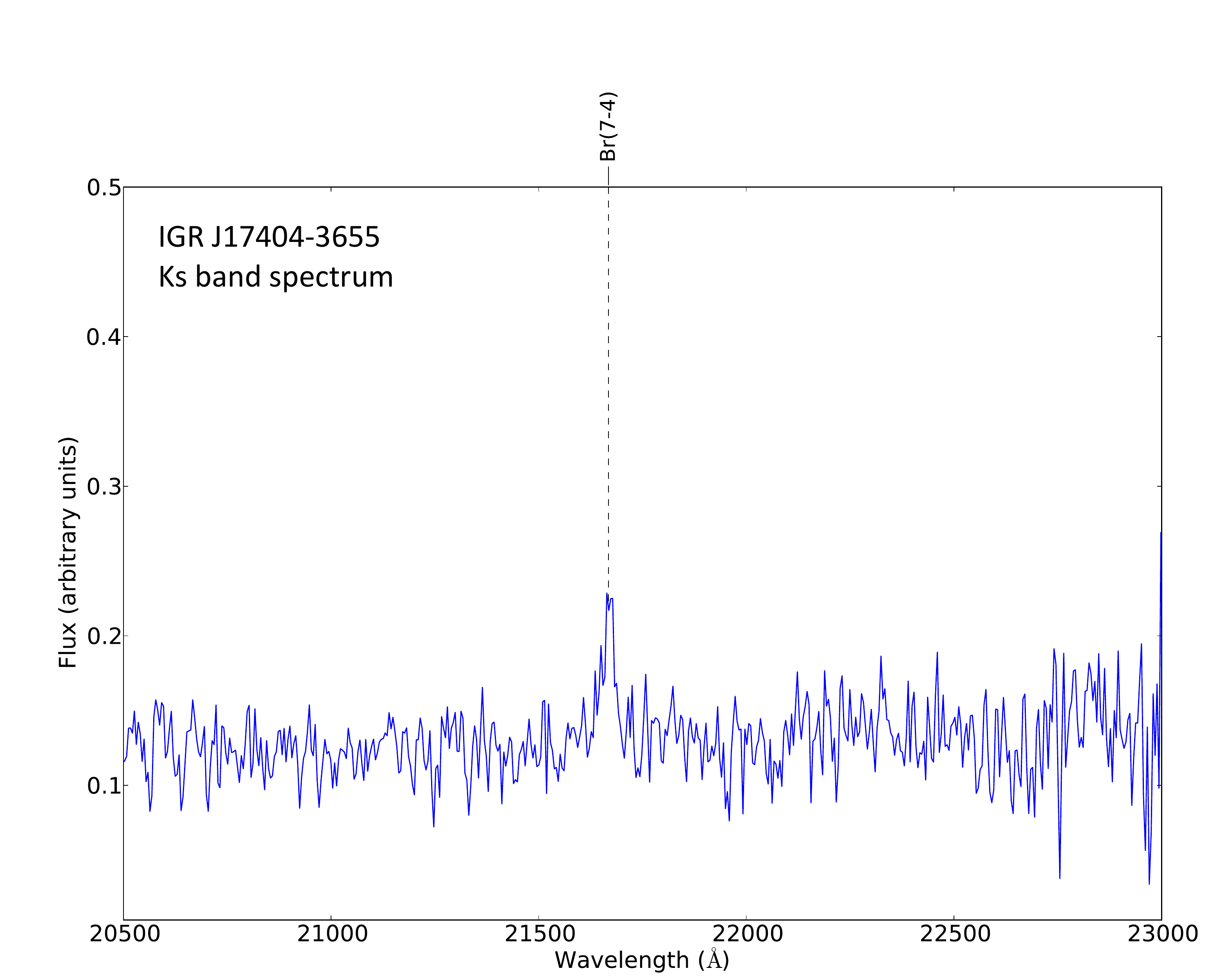}\\
\includegraphics[width=0.5\textwidth]{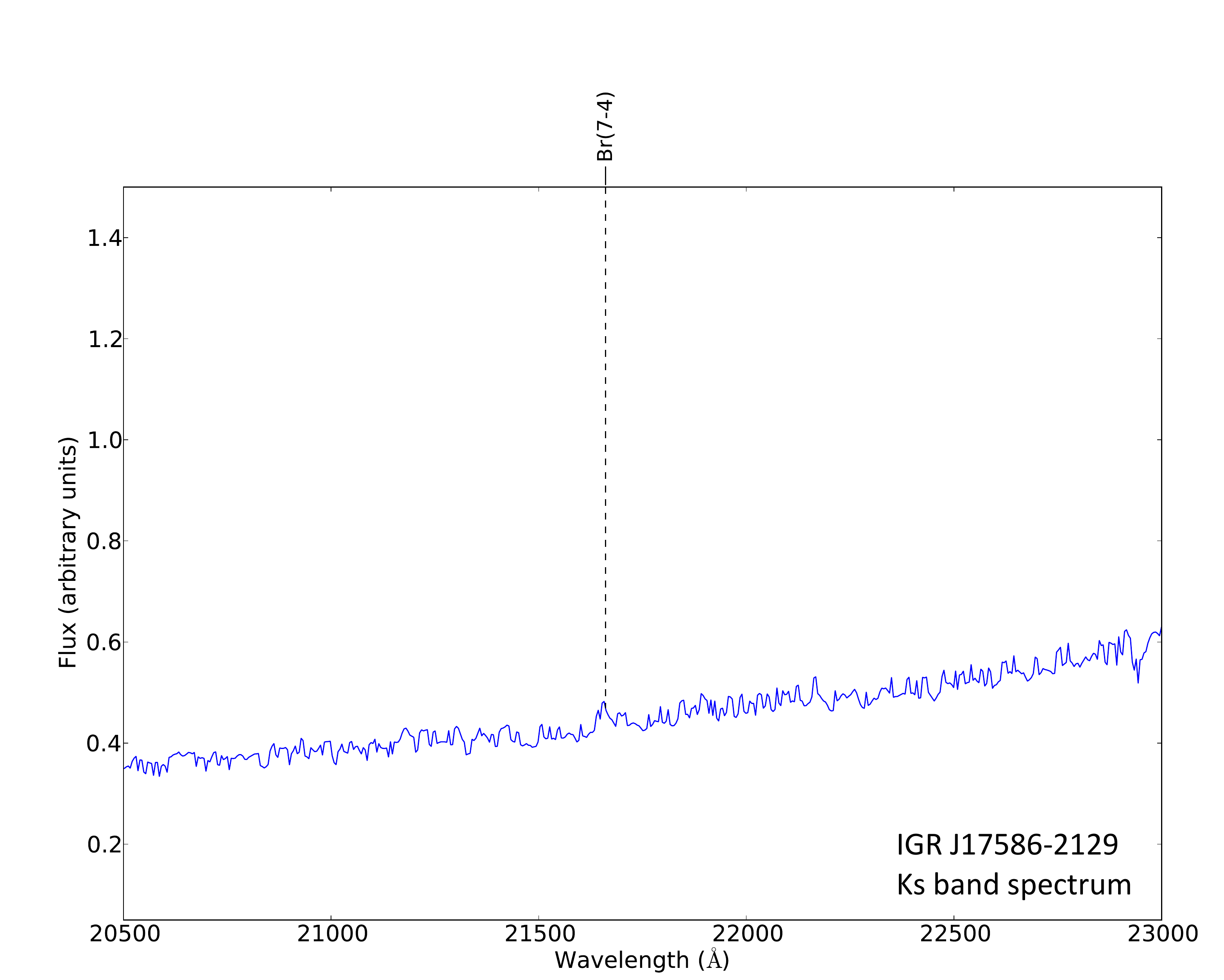}\hfill
\end{center}
\end{minipage}
\caption{Spectra of IGR J17354-3255 (top left), IGR J17404-3655 (top right), and IGR J17586-2129 (bottom).} 
\label{fourth}
\end{figure*}

\begin{figure*}[h!]
\begin{minipage}[t]{1\textwidth}
\begin{center}
\includegraphics[width=0.5\textwidth]{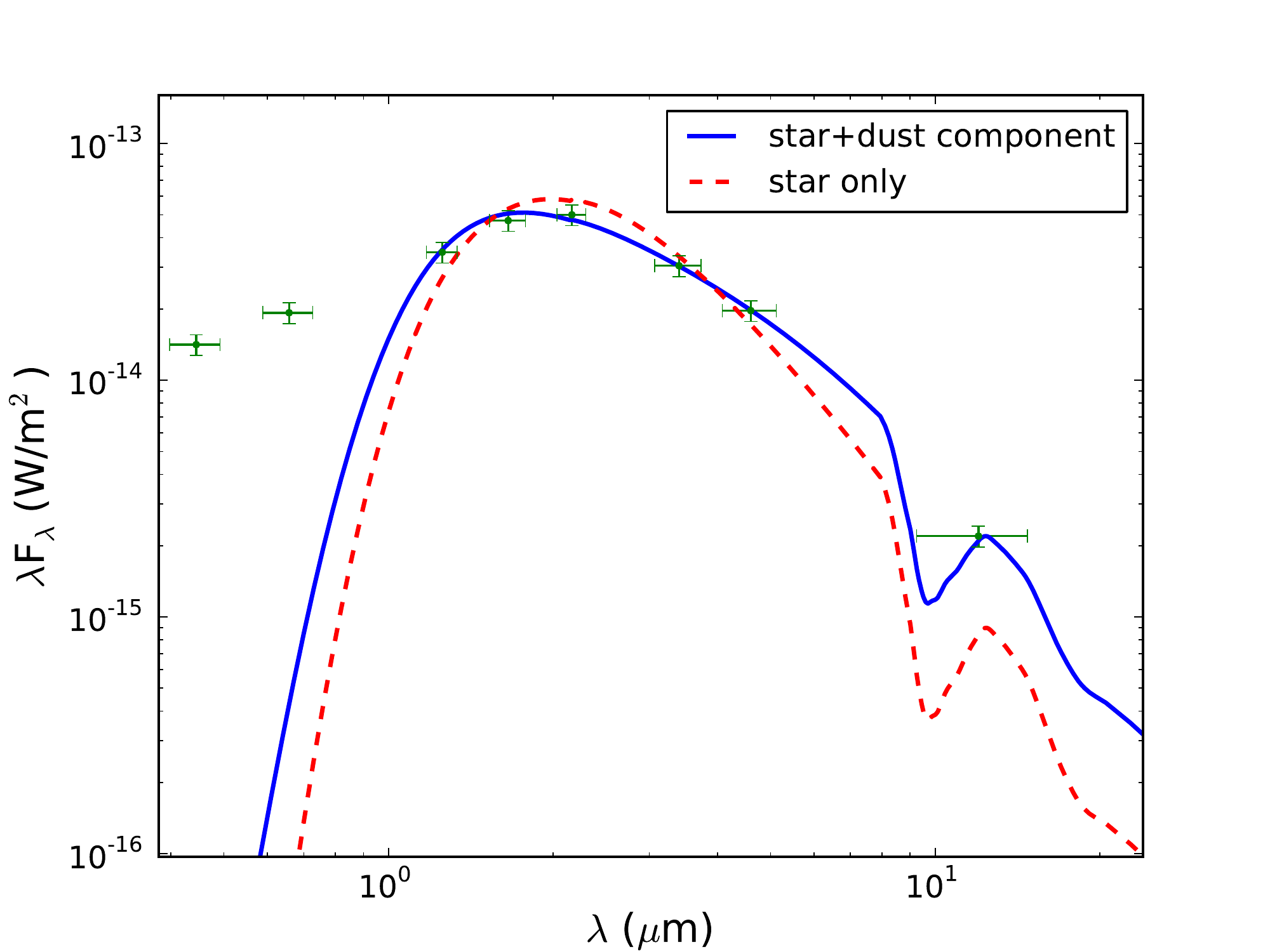}\hfill
\includegraphics[width=0.5\textwidth]{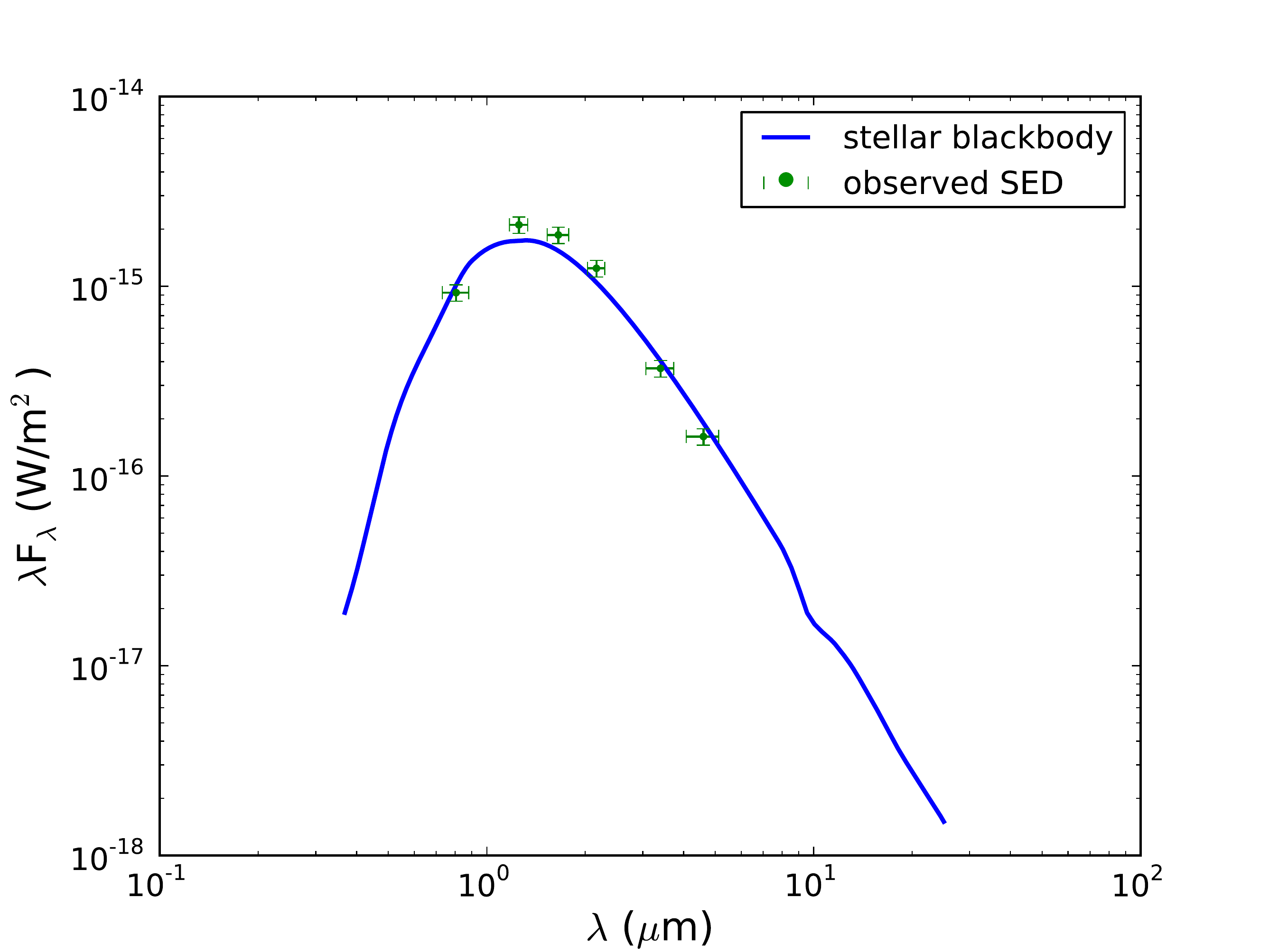}\\
\includegraphics[width=0.5\textwidth]{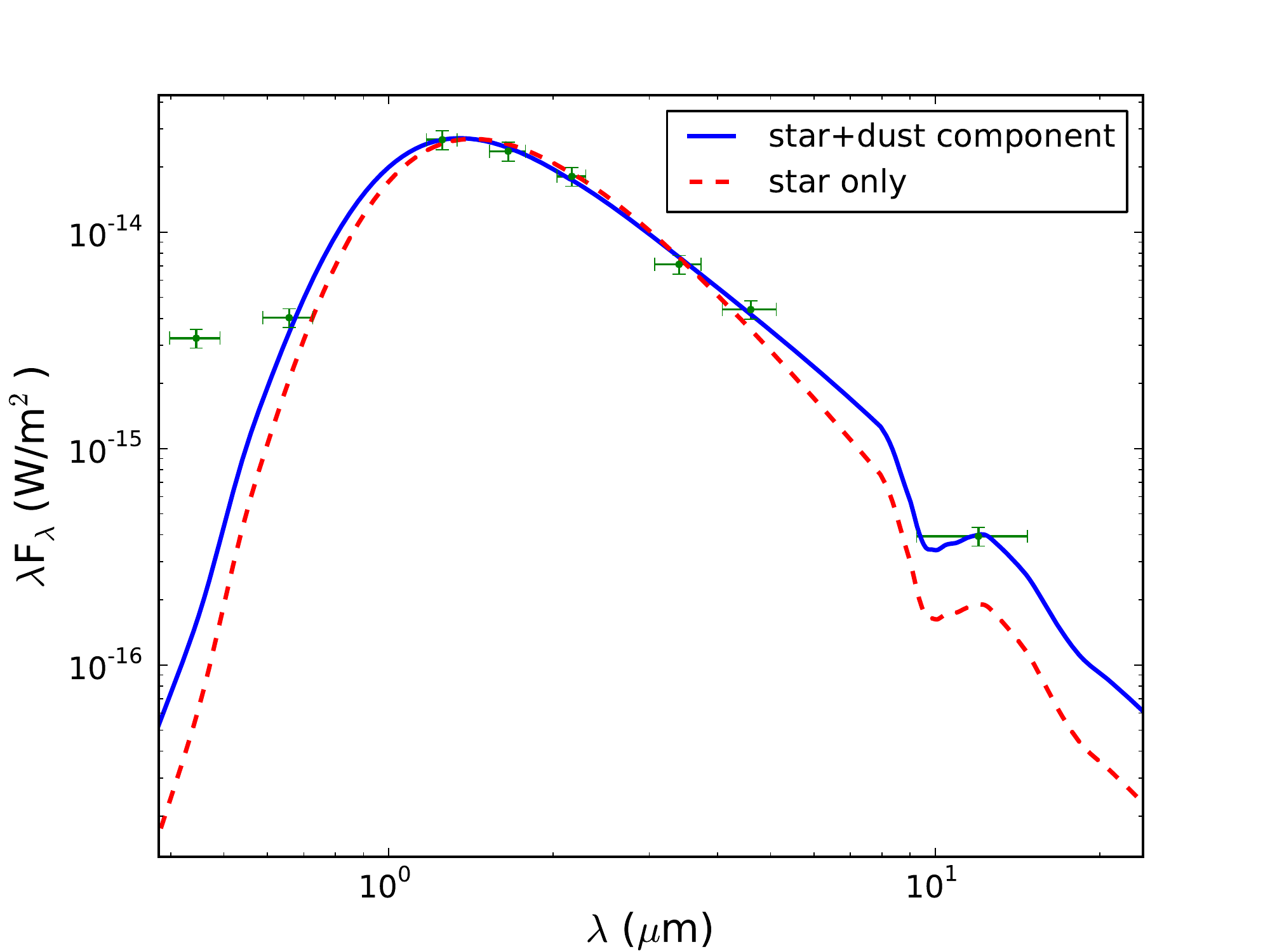}\hfill
\includegraphics[width=0.5\textwidth]{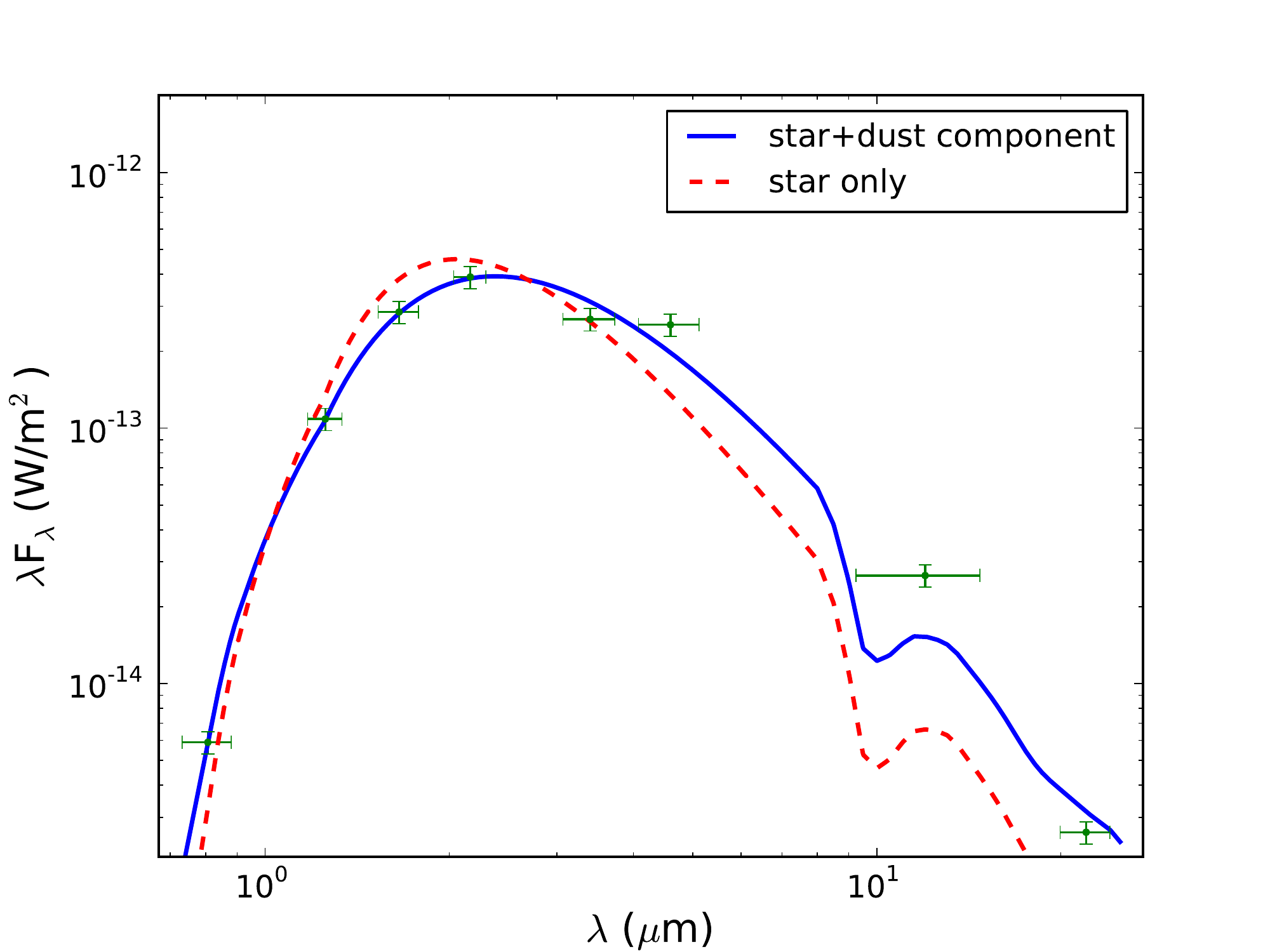} \\

\end{center}
\end{minipage}
\caption{Best fit of IGR J10101-5654 (top left), IGR J11187-5438 (top right), IGR 11435-6109 (bottom left), and IGR J17586-2129 (bottom right). Derived parameters are given in Table \ref{SED_param_fit}.}
\label{best_fit}
\end{figure*}

Only the Br(7-4) line is detected in emission in our nIR $K_s$-band spectrum (see Figure \ref{fourth}). Moreover, the SED (I, J, H, $K_S$, 3.4 $\mu$m, 4.6 $\mu$m, 12 $\mu$m, 22 $\mu$m photometry points) is well fitted ($\chi^2$/dof= 5.0, see Figure \ref{best_fit}) with a model that combines two absorbed blackbodies: one representing the companion star emission, and another one representing a possible mIR excess due to absorbing material that enshrouds the companion star \citep[see e.g.][]{Rahoui2008}.  The free parameters of the fits are the absorption in V band, $A_V$, the radius to distance ratio $R_*/D$, the additional blackbody component temperature $T_d$ and its radius $R_d$. The stellar blackbody temperature was fixed to 20000 K. The best-fitting parameters are stellar radius over distance $R_*/D$ = 5.47 $\pm$ 0.27 $R_{\sun}$/kpc , along with a dust component: dust temperature $T_d$ = 2033 $\pm$ 304 K, and spatial extension $R_d/D$ = 49.01 $\pm$ 7.35 $R_{\sun}$/kpc. The retrieved extinction value is equal to $A_V$ = 13.6 $\pm$ 0.68 mag. For comparison, the best-fitting parameters without the additional component are $A_V$ =17.0 $\pm$ 0.34 mag; $R/D$ = 10.53 $\pm$ 0.21 $R_{\sun}$/kpc, and $\chi^2$/dof = 24.14. Thus, the SED fitting points toward a supergiant companion star.

%\begin{figure}[h!]
%\begin{center}
%\includegraphics[scale=0.45]{fit_17586.pdf}
%\caption{Best fit of IGR J17586-2129. Derived parameters: $A_V$ = 13.6 mag; $T_*$ = 20000 K; $R/D$ = 10.53 $\pm$ 0.21 $R_{\sun}$/kpc; $T_d$ = 2033 K; $R_d$ = 49.01$R_{\sun}$/kpc; $\chi^2_{min} = 5.0$.}
%\label{fit17586}
%\end{center}
%\end{figure}

%\begin{figure*}[th!]
%\begin{center}
%\includegraphics[scale=0.25]{fourthgroup.pdf}
%\caption{Spectra of IGR J17354-3255, IGR J17404-3655, IGR J17586-2129. }
%\label{fourth}
%\end{center}
%\end{figure*}

%------------------------------------------------------------------------------------------------------------------------------------------------------------------------------------

\begin{center}
\begin{table*}
\scalebox{0.9}
{
\begin{tabular}{ccccccc}
  \hline
  \hline
  Source & Line & $\lambda$ ($\mu$m) & $\lambda_{fit}$ ($\mu$m) & Flux (arbitrary units) & EW (\AA) & FWHM (\AA)\\
  \hline
  \textbf{IGR J10101-5654} & & & & & &\\
        \multicolumn{7}{c}{\textbf{$H$ band}}\\
&Br(20-4)&1.519 & 1.5198 & 455.8 & -1.383 & 19.87\\
&Br(19-4)&1.525 & 1.5265 & 93.22 & -0.279 & 4.304\\
&Br(18-4)&1.534 & 1.5354 & 555.4 & -1.662 & 23.57\\
&Br(17-4)&1.543 & 1.5451 & 634.1 & -1.905 & 24.53\\
&Br(16-4)&1.555 & 1.5568 & 901.8 & -2.692 & 24.88\\
&Br(15-4)&1.569 & 1.5708 & 707.6 & -2.031 & 25.34\\
&\ion{N}{iii} + \ion{C}{iii}  &1.575 & 1.5778 & 841.4 & -2.439 & 44.81\\
&+ \ion{Fe}{ii}  & 1.576 & & & &\\
&Br(14-4)&1.588 & 1.5901 & 1481 & -4.134 & 32.46\\
&Br(13-4)&1.611 & 1.6120 & 1151 & -3.167 & 31.88\\
&Br(12-4)&1.641 & 1.6410 & 634.9 & -1.649 & 15.93\\
&$\left[\rm{\ion{Fe}{ii}}\right]$  & 1.643 & 1.6427 & 572 & -1.488 &  12.58\\
%& & 1.6717 & -430.6 & 1.033 & 18.28\\
%& & 1.6746 & -214.8 & 0.5135 & 8.593\\
&Br(11-4)& 1.681 & 1.6822 & 1121 & -2.645 & 25.33\\
&\ion{Fe}{ii} & 1.688 & 1.6893 & 154.5 & -0.363 & 7.551\\
&$\left[\rm{\ion{Fe}{ii}}\right]$ &1.711 & 1.7122 & 362.5 & -0.824 & 15.66\\
&Br(10-4)& 1.736 & 1.7377 & 1032 & -2.062 & 24.4\\
   \multicolumn{7}{c}{\textbf{$K_s$ band}}\\
  &\ion{He}{i} & 2.058 & 2.0586 & 1537 & -4.47 & 57.47\\
  &H$_2$ & 2.1218&2.1226&269.1&-0.491&10.59\\
  % & & 2.1065 & 503.5 & -1.438 & 45.89\\
  % & & 2.1225 & 169.5 & -0.473 & 13.21\\
  & \ion{Mg}{ii}  & 2.138 & 2.1363 & 326.9 & -0.901 & 35.5\\
   &\ion{Mg}{ii} & 2.144 & 2.1454 & 344.9 & -0.941 & 24.12\\
   &Br(7-4) & 2.166 & 2.1659 & 2249 & -5.952 & 29.98\\
   &\ion{He}{ii} & 2.1891 & 2.1886 & 334.1 & -0.872 & 20.2\\
   &\ion{Na}{i} & 2.206 & 2.2064 & 328.2 & -0.833 & 15.24\\
  % FeI & 2.2387 & 2.2388 & 168.5 & -0.416 & 9.509\\
  % & & 2.2428 & 327.1 & -0.808 & 30.18\\
   %& & 2.2480 & 138.7 & -0.339 & 11.6\\
   %& & 2.2528 & 162.2 & -0.391 & 12.72\\
   %& & 2.2616 & 313.5 & -0.739 & 16.73\\
   %MgI & 2.2814 & 2.2811 & 258.8 & -0.594 & 10.38\\
   %& & 2.2828 & 214.1 & -0.496 & 10.69\\
   %& & 2.2891 & 159.8 & -0.363 & 6.285\\
   \hline
   
   %-------------------------------------------------------------------------------------------------------------------
     \textbf{IGR J11435-6109} & & & & & &\\

   \multicolumn{7}{c}{\textbf{$H$ band}}\\
&Br(21-4) & 1.513 & 1.5137 & 310.8 & -1.285 & 24.42\\
&Br(20-4) &1.519 & 1.5200 & 316.3 & -1.323 & 23.2\\
&Br(19-4)&1.525 & 1.5260 & 512.6 & -2.178 & 31.92\\
&Br(18-4)&1.534 & 1.5349 & 652.3 & -2.799 & 26.48\\
&Br(17-4)&1.543 & 1.5441 & 571.6 & -2.458 & 28.05\\
&Br(16-4) & 1.555& 1.5568 & 472.2 & -1.993 & 28.01\\
&Br(15-4) & 1.569 & 1.5707 & 528.9 & -2.205 & 24.76\\
&\ion{N}{iii} + \ion{C}{iii}  &1.575 & 1.5772 & 344.1 & -1.43 & 22.46\\
&+ \ion{Fe}{ii}  & 1.576 & & & &\\
&Br(14-4)&1.588 & 1.5895 & 1013 & -4.158 & 31.53\\
&Br(13-4)&1.611 & 1.6119 & 957.2 & -3.896 & 43.94\\
&Br(12-4)&1.641 & 1.6414 & 894.7 & -3.508 & 38.21\\
&Br(11-4)&1.681 & 1.6813 & 1032 & -3.8 & 36.15\\
&\ion{He}{i}  & 1.700 & 1.7014 & 53.19 & -0.19 & 4.389\\
&$\left[\rm{\ion{Fe}{ii}}\right]$ ?& 1.711 & 1.7120 & 306.2 & -1.101 & 20.43\\
&Br(10-4)&1.736 & 1.7364 & 973.4 & -3.225 & 27.62\\

    \multicolumn{7}{c}{\textbf{$K_s$ band}}\\
    
&\ion{He}{i} &2.058 & 2.0590 & 384.9 & -2.788 & 33.86\\
% & & 2.0915 & 49.13 & -0.357 & 6.496\\
% & & 2.1082 & 126.4 & -0.923 & 13.15\\
% & & 2.1225 & 215.6 & -1.529 & 43.63\\
& \ion{Mg}{ii} & 2.138 & 2.1385 & 193.8 & -1.367 & 33.36\\
 &\ion{Mg}{ii} & 2.144 & 2.1444 & 91.25 & -0.637 & 8.416\\
 %& & 2.1466 & 124 & -0.867 & 19.22\\
 &Br(7-4) & 2.166 & 2.1664 & 273.9 & -1.805 & 16.35\\
 %& & 2.1851 & 227.7 & -1.523 & 35.55\\
 &\ion{He}{ii} & 2.1891 & 2.1893 & 125.7 & -0.83 & 20.09\\
   \hline
 %-------------------------------------------------------------------------------------------------------------------
     \textbf{IGR J13020-6359} & & & & & &\\
 \multicolumn{7}{c}{\textbf{$K_s$ band}}\\
&\ion{He}{i} & 2.058 & 2.0594 & 714.3 & -2.734 & 25.74\\
%& & 2.0929 & 158.5 & -0.619 & 11.12\\
%& & 2.0974 & 419.9 & -1.671 & 20.78\\
&Br(7-4) & 2.166 & 2.1663 & 2962 & -9.986 & 29.09\\
%& &  2.2116 & 256 & -0.869 & 12.7\\
  \hline
%-------------------------------------------------------------------------------------------------------------------
     \textbf{IGR J14331-6112} & & & & & &\\
 \multicolumn{7}{c}{\textbf{$K_s$ band}}\\
&Br(7-4) & 2.166 & 2.1657 & 184.3 & -4.148 & 34.54\\
   \hline
%-------------------------------------------------------------------------------------------------------------------
     \textbf{IGR J14488-5942} & & & & & &\\
       \multicolumn{7}{c}{\textbf{$K_s$ band}}\\
   & \ion{He}{i} & 2.058 & 2.0590 & 446.6 & -9.32 & 39.86\\
  %  \ion{C}{iv} ? & 2.078 & 2.0788 & 236.7 & -4.663 & 33.97\\
    %& & 2.1174 & -60.71 & 1.108 & 13.62\\
    %& & 2.1494 & -52.28 & 0.8738 & 4.307\\
    %& & 2.1510 & -42.55 & -0.6936 & 2.484\\
    %& & 2.1530 & -129.5 & 2.112 & 16.17\\
    %& & 2.1597 & 72.1 & -1.186 & 8.545\\
   % & & 2.1616 & 77.61 & -1.263 & 7.775\\
    &Br(7-4) & 2.166 & 2.1665 & 357.5 & -5.722 & 35.7\\
    %& & 2.1764 & -54.16 & 0.8593 & 4.218\\
    %& & 2.1822 & -86.09 & 1.335 & 8.958\\
    %& & 2.2016 & -69.27 & 1.07 & 4.322\\
    %& & 2.2037 & -103.1 & 1.566 & 13.95\\
    %& & 2.2196 & 90.99 & -1.406 & 4.393\\
    %& & 2.2280 & 196.4 & -3.034 & 15.46\\
    %& & 2.2344 & -113.9 & 1.638 & 12.14\\
    %& & 2.2427 & -79.71 & 1.147 & 3.498\\
    %& & 2.2516 & 111.8 & -1.618 & 10.59\\
    %& & 2.2525 & -81.63 & 1.155 & 4.132\\
  %  $\left[\rm{\ion{Fe}{ii}}\right]$ ? & 2.254 & 2.2539 & 77.22 & -1.099 & 9.008\\
   % & & 2.2551 & 32.64 & -0.47 & 2.264\\
   % & & 2.2575 & -50.58 & 0.719 & 2.97\\
   % & & 2.2746 & -72.13 & 1.023 & 4.367\\
   % MgI & 2.2814 & 2.2814 & -84.56 & 1.168 & 5.334\\ %in absorption ???
   % & & 2.2944 & 139.4 & -1.981 & 6.798\\
   % & & 2.2974 & -113.7 & 1.589 & 3.758\\
      \hline
%-------------------------------------------------------------------------------------------------------------------
     \textbf{IGR J16195-4945} & & & & & &\\
      \multicolumn{7}{c}{\textbf{$H$ band}}\\
  &Br(19-4) & 1.525 & 1.5251 & 113.2 & -1.134 & 7.221\\
  &Br(18-4) & 1.534 & 1.5339 & 177.4 & --1.808 & 11.44\\
  %&& & 1.5397 & 159 & -1.615 & 13.92\\
 % && & 1.5461 & 155 & -1.591 & 11.57\\
  &Br(16-4)&1.555& 1.5566 & 137.5 & -1.422 & 10.83\\
 % && & 1.5608 & 95.9 & -0.961 & 6.802\\
 &$\left[\rm{\ion{Fe}{ii}}\right]$ ?& 1.599 & 1.5975 & 337.4 & -3.148 & 16.79\\
 % && & 1.6033 & 80.26 & -0.694 & 3.939\\
 %&& & 1.608 & -75.6 & 0.6205 & 6.839\\
 &Br(13-4)&1.611 & 1.6129 & 213.4 & -1.852 & 14.22\\
 %&& & 1.6234 & 298.2 & -2.349 & 22.03\\
 & \ion{He}{i} & 1.700 & 1.6998 & -731.6 & 3.87 & 41.00\\
 %tel_16195_H_2 est peut-etre mieux au niveau de la correction tellurique....
   \multicolumn{7}{c}{\textbf{$K_s$ band}}\\
    &\ion{He}{i} & 2.058 & 2.0577 & -359.4 & 0.9498 & 11.37\\
    &\ion{He}{i} & 2.112 & 2.1120 & -534.9 & 1.367 & 16.17\\
    & \ion{He}{i} & 2.150 & 2.1510 & -400.4 & 1.008 & 18.73\\
    &Br(7-4) & 2.166 & 2.1672 & 1629 & -4.005 & 37.3\\
    &\ion{Na}{i}  & 2.209 & 2.2091 & 497 & -1.129 & 30.87\\
    && & 2.2218 & -353 & 0.7856 & 10.02\\
    %& & 2.2301 & 231.1 &  -0.51 & 6.136\\
  %  & & 2.2486 & 340.8 & -0.728 & 11.03\\
   % & & 2.2802 & -695.4 & 1.271 & 18.38\\
       \hline
  \hline
  \end{tabular}
}
  \caption{Spectroscopy results for all the sources (cont. Table \ref{lines_all_2}). We indicate the identification of the lines, the rest wavelength ($\mu$m), the fitted central wavelength ($\mu$m), the flux (in arbitrary units), the equivalent width (EQW in \AA, negative and positive values indicate emission and absorption lines, respectively) and the FWHM in \AA.}
    \label{lines_all}
\end{table*}
\end{center}

\begin{center}
\begin{table*}
\scalebox{0.8}
{
\begin{tabular}{ccccccc}
  \hline
  \hline
  Source & Line & $\lambda$ ($\mu$m) & $\lambda_{fit}$ ($\mu$m) & Flux (arbitrary units) & EW (\AA) & FWHM (\AA)\\
  \hline

     \textbf{IGR J16318-4848} & & & & & &\\
  \multicolumn{7}{c}{\textbf{$K_s$ band}}\\
 &$\left[\rm{\ion{Fe}{ii}}\right]$&2.046 & 2.0451 & 4853 & -0.427 & 20.54\\
 &\ion{He}{i} & 2.058& 2.0590 & 470387 & -38.38 & 29.34\\
 &\ion{Fe}{ii} &2.089 & 2.0893 & 127073 & -10.89 & 36.01\\
 && & 2.0969 & 7202 & -0.618 & 21.75\\
 &\ion{He}{i} & 2.1126 & 2.1135 & 45194 & -3.756 & 25.47\\
 &\ion{N}{iii} + \ion{C}{iii} &2.116 & 2.1166 & 22009 & -1.813 & 30.62\\
 & \ion{H$_2$}{} ? &2.1218 & 2.1218 & 7961 & -0.646 & 30.69\\
&\ion{Mg}{ii} &2.138 & 2.1372 & 78531 & -6.139 & 30.58\\
&\ion{Mg}{ii}& 2.144& 2.1440 & 42806 & -3.293 & 32.4\\
&Br(7-4)&2.1661 & 2.1662 & 729029 & -52.55 & 32.28\\  %lŽgre bosse dans l'aile gauche
&\ion{He}{i} & 2.1847& 2.1854 & 19165 & -1.411 & 30.17\\
&\ion{He}{ii} & 2.189 & 2.1886 & 8241 & -0.599 & 18.66\\
&\ion{Na}{i} ? &2.2056 & 2.2067 & 48686 & -3.526 & 31.06\\
&\ion{Na}{i} & 2.209 & 2.2097 & 18470 & -1.337 & 30.35\\
&$\left[\rm{\ion{Fe}{ii}}\right]$ ? &2.224 & 2.2263 & 27940 & -1.974 & 64.4\\%peut-tre deux raies blended...
&\ion{Fe}{ii} & 2.240 &2.2409 & 26749 & -1.827 & 41.35\\
&& & 2.2818 & 16748 & -1.051 & 17.12\\
&& & 2.2836 & 7622 & -0.474 & 15.83\\
  \hline
%-------------------------------------------------------------------------------------------------------------------
     \textbf{IGR J16320-4751} & & & & & &\\
  \multicolumn{7}{c}{\textbf{$K_s$ band}}\\
  &\ion{He}{i} & 2.058 & 2.0586 & 711.3 & -1.719 & 21.02\\
%& & 2.0649 & -59.49 & 0.1393 & 3.392\\
%& & 2.0679 & -92.65 & 0.2165 & 4.658\\
%& & 2.0706 & -87.38 & 0.2019 & 2.164\\
%& & 2.0748 & -131.7 & 0.3049 & 4.803\\
& \ion{C}{iv} ? & 2.078 & 2.0790 & -66.08 & 0.1536 & 2.884\\
%& & 2.0845 & -133.3 & 0.3067 & 11.26\\
%& & 2.0916 & -102.3 & 0.2321 & 3.45\\
%& & 2.0945 & -95.84 & 0.214 & 4.176\\
%& & 2.1054 & -182.5 & -0.407 & 13.46\\
%& & 2.1074 & 554.4 & -1.217 & 25\\
&\ion{He}{i} &2.112 &2.1127 & -251.6 & 0.5491 & 6.609\\
%& & 2.1172 & -148.2 & 0.3081 & 6.623\\
%& & 2.1202 & -177.1 & 0.3666 & 13.63\\
%& & 2.1248 & -244.8 & 0.5078 & 18.34\\
%& &2.1296 & 84.91 & -0.177 & 10.02\\
%& & 2.1324 & -66.56 & 0.1382 & 1.992\\
&\ion{Mg}{ii}  & 2.138& 2.1371 & 590.4 & -1.229 & 26.38\\
%& & 2.1558 & -115.9 & 0.2244 & 3.463\\
&Br(7-4)& 2.1661& 2.1668 & 2587 & -4.965 & 37.95\\
%& & 2.1872 & -148.9 & 0.2666 & 3.406\\
%& & 2.2031 & -182.2 & 0.3196 & 5.906\\
%& & 2.2133 & -144 & 0.248 & 3.439\\
%& & 2.2153 & -164.3 & 0.2835 & 4.822\\
%& & 2.2224 & -237.8& 0.4012 & 7.172\\
%& & 2.2289 & -71.22 & 0.1183 & 2.743\\
%& & 2.2522 & -178.9 & 0.2798 & 5.711\\
%& & 2.2635 & 525.1 & -0.803 & 18.06\\
%& & 2.2892 & -789 & 1.12 & 13.94\\
       \hline
%-------------------------------------------------------------------------------------------------------------------
     \textbf{IGR J16328-4726} & & & & & &\\
  \multicolumn{7}{c}{\textbf{$K_s$ band}}\\
&\ion{He}{i} & 2.058 & 2.0579 & -552.6 & 1.844 & 17.11\\
%& & 2.0627 & 89.36 & -0.296 & 6.159\\
& \ion{C}{iv} & 2.069 & 2.0687 & 134 & -0.438 & 7.924\\
&\ion{He}{i}  & 2.112 - 2.113  & 2.1114 & -234.1 & 0.7032 & 15.94\\
&\ion{N}{iii}/\ion{C}{iii} & 2.1155 & 2.1155 & 896.4 & -2.662 & 33.52 \\
&Br(7-4)& 2.1661& 2.1661 & 4195 & -11.61 & 55.84\\
&\ion{He}{ii} & 2.189& 2.1898 & -173.4 & 0.4762 & 11.37\\
&\ion{Na}{i}  &2.206  & 2.2072 & 589.9 & -1.591 & 40.87\\
  \hline
  %-------------------------------------------------------------------------------------------------------------------
     \textbf{IGR J16418-4532} & & & & & &\\
  \multicolumn{7}{c}{\textbf{$K_s$ band}}\\
   & \ion{He}{i} &2.058 & 2.0580 & 340.2 & -1.079 & 50.27\\
    &\ion{He}{i} & 2.112 & 2.1124 & -337.6 & 1.017 & 18.15\\
    &Br(7-4) & 2.1661 & 2.1672 & 1470 & -4.233 & 53.47\\
       \hline
  %-------------------------------------------------------------------------------------------------------------------
     \textbf{IGR J17354-3255} & & & & & &\\      
\multicolumn{7}{c}{\textbf{$K_s$ band}}\\
   &\ion{He}{i} &2.058& 2.0614 & -770 & 0.767 & 18.01\\
   &\ion{He}{i} & 2.112 & 2.1138 & -1549 & 1.524 & 18.87\\
   &\ion{Na}{i} & 2.206 & 2.2065 & 698.9 & -0.635 & 14.06\\
   &\ion{Na}{i} & 2.209 & 2.2080 & 1639 & -1.483& 33.41\\
       \hline
        %-------------------------------------------------------------------------------------------------------------------
     \textbf{IGR J17404-3655} & & & & & &\\      
 \multicolumn{7}{c}{\textbf{$K_s$ band}}\\

  & Br(7-4 )& 2.1661 & 2.1668 & 581.2 & -26.67 & 36.44\\
 \hline
   %-------------------------------------------------------------------------------------------------------------------
     \textbf{IGR J17586-2129} & & & & & &\\      
 \multicolumn{7}{c}{\textbf{$K_s$ band}}\\
&Br(7-4)& 2.1661 & 2.1659 & 1231 & -5.965 & 46.87\\
%& & 2.2165 & 405.2 & -1.715 & 13.9\\
%& & 2.2943 & -974.7 & 3.222 & 27.02\\
       \hline
  \hline
\end{tabular}
}
  \caption{Spectroscopy results for all the sources (cont.). We indicate the identification of the lines, the rest wavelength ($\mu$m), the fitted central wavelength ($\mu$m), the flux (in arbitrary units), the equivalent width (EQW in \AA, negative and positive values indicate emission and absorption lines respectively) and the FWHM in \AA.}
    \label{lines_all_2}
\end{table*}
\end{center}

\begin{table*}
\begin{tabular}{llllll}
  \hline
  \hline
  Line & $\lambda$ ($\mu$m) & EW ($\AA$) -- 2010 data & EW ($\AA$) -- 2003 data& FWHM ($\AA$) -- 2010 data  & FWHM ($\AA$) -- 2003 data\\
    \hline 
$\left[\rm{\ion{Fe}{ii}}\right]$&2.046 & -0.427 & -1 & 20.54  & 26 \\
 \ion{He}{i} & 2.058 & -38.38 & -42 & 29.34 & 35 \\
 \ion{Fe}{ii} &2.089 & -10.89 &-13 & 36.01 & 36\\
% & & 2.0969 & 7202 & -0.618 & 21.75\\
 \ion{He}{i} & 2.1126 & -3.756&-5 & 25.47&43\\
 \ion{N}{iii} + \ion{C}{iii} &2.116  & -1.813 & -4 (?) & 30.62 & 44 (?)\\
 %& & 2.1218 & 7961 & -0.646 & 30.69\\
\ion{Mg}{ii} &2.138 &  -6.139 & -8& 30.58&35\\
%\ion{Mg}{ii}& 2.144& 2.1440 & 42806 & -3.293 & 32.4\\
Br(7-4)&2.1661  & -52.55 & -45& 32.28 & 36\\  %lŽgre bosse dans l'aile gauche
\ion{He}{i} & 2.1847 & -1.411 &-7 (?)& 30.17&50\\
%\ion{He}{ii} & 2.189 & 2.1886 & 8241 & -0.599 & 18.66\\
\ion{Na}{i} ? &2.2056 & -3.526 & -3 & 31.06 & 40\\
\ion{Na}{i} & 2.209 & -1.337 & blended &30.35&blended  \\
$\left[\rm{\ion{Fe}{ii}}\right]$  &2.224 & -1.974 & -0 & 64.4 & 21\\%peut-tre deux raies blended...
\ion{Fe}{ii} & 2.240  & -1.827 & -2 & 41.35 & 40\\
%& & 2.2818 & 16748 & -1.051 & 17.12\\
%& & 2.2836 & 7622 & -0.474 & 15.83\\
   \hline
\end{tabular}
  \caption{Equivalent width and FWHM values of IGR J16318-4848. Comparison of our data with those of \citet{Filliatre2004}. No major variation are detected for the brightest lines. Other lines exhibit variations between 25$\%$ and 65$\%$, but their faintness prevents any further conclusion on their variabilities.}
  \label{16318_var}
\end{table*}

\begin{table*}
\begin{tabular}{lccccccc}
     \hline
  \hline
  Source & $A_V$ & $T_*$ & $R/D$ & $T_d$ & $R_d/D$ & $\chi^2/dof$ & $dof$\\
  & (mag) & (K) & ($R_{\sun}$/kpc) & (K) & ($R_{\sun}$/kpc) & &\\ 
  \hline
  IGR J10101-5654 & 13.1 $\pm$ 0.86 & 20000 & 2.79 $\pm$ 0.20 & 1016 $\pm$ 127 & 28.3 $\pm$ 4.25 & 2.23 & 2\\
  IGR J11187-5438 & 3.9 $\pm$ 0.40 & 6000 & 0.60 $\pm$ 0.03 &  -- & -- & 3.45 & 3\\
  IGR J11435-6109 & 7.98 $\pm$ 0.44 & 20000 & 1.33 $\pm$ 0.050 & 925.6 $\pm$ 190.7 & 51.6 $\pm$ 2.97 & 1.14 & 2\\
  IGR J17586-2129 &   13.6 $\pm$ 0.68 & 20000 & 5.47 $\pm$ 0.27 & 2033 $\pm$ 3045 & 49.01 $\pm$ 7.35 & 5.0 & 4\\ 
     \hline
\end{tabular}
  \caption{Parameters derived from SED fitting.}
  \label{SED_param_fit}
\end{table*}

\subsection{Search for counterparts and study of the environment with \textit{\textit{WISE}} data}

\begin{figure}[th!]
\begin{center}
\includegraphics[scale=0.35]{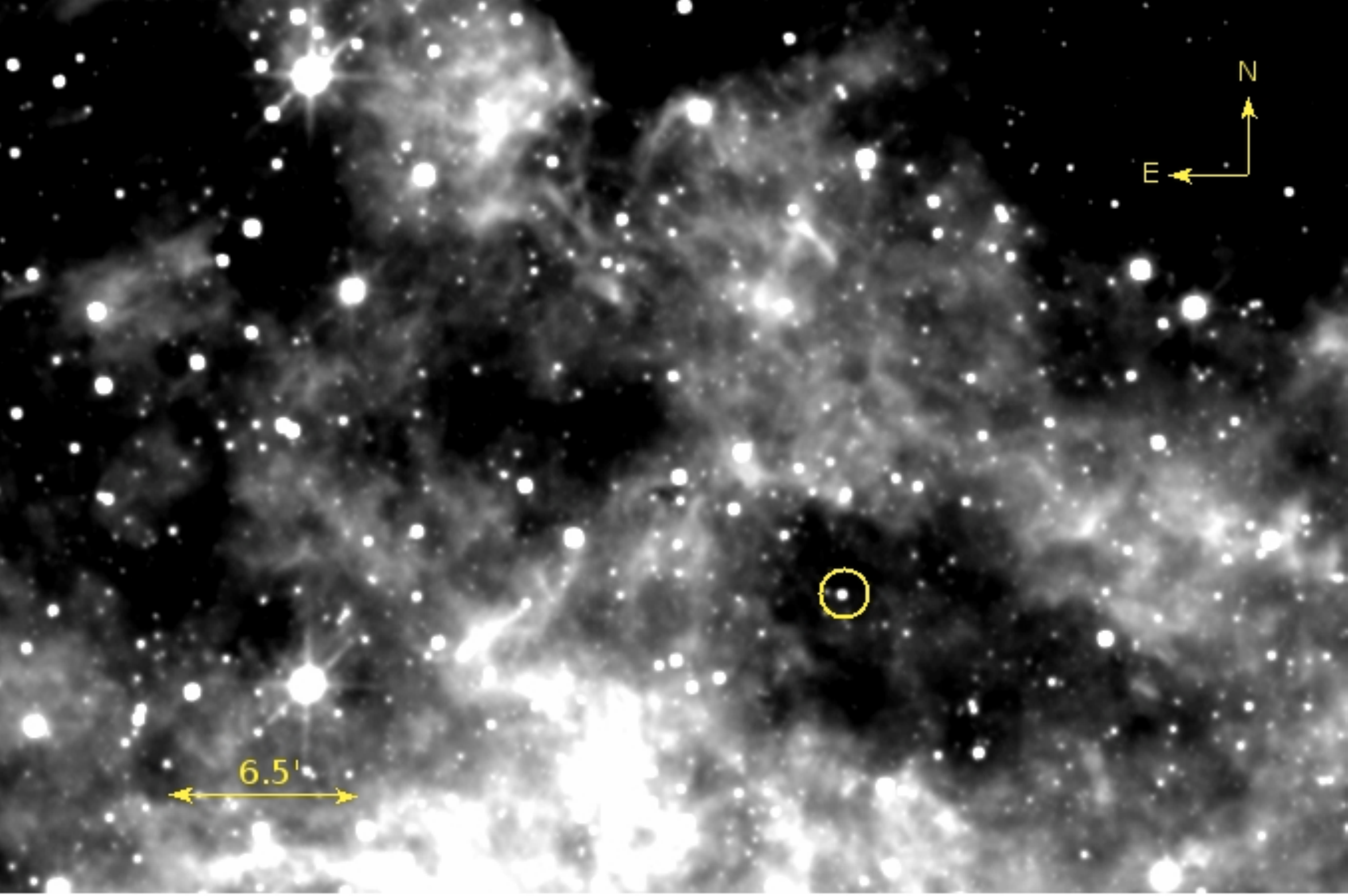}
\caption{\textit{WISE} 12 $\mu$m image of the region around IGR J11435-6109.}
\label{11435_WISE}
\end{center}
\end{figure}

With the aim of studying the close environment of HMXB, we carried out a systematic search for counterparts in the \textit{\textit{WISE}} or GLIMPSE (when the source is blended) catalogs . This allows us to detect a mIR excess around the HMXB due to circumbinary dust (see the discussion for IGR J10101-5654, IGR J11435-6109, IGR J17586-2129), and examine the larger scale structures that probe the feedback of the source on its environment as well as the impact of the environment on the HMXB (see Figure \ref{11435_WISE}).
%Any quantitative analyze is out of the scope of this paper and sources presenting possible interaction with its close environment will be the subject of a forthcoming paper.

We present in Table \ref{WISE} the fluxes in the four WISE filters for each detected source. Fluxes $F_\nu$ (in Jansky) were derived from the Vega magnitude $m_{\mathrm{Vega}}$, using the formula
\begin{equation}
F_\nu [Jy] = F_{\nu0}\times10^{(-m_{\mathrm{Vega}}/2.5)},
\end{equation}
where $F_{\nu0}$ is the zero-magnitude flux density derived for sources with power-law spectra $F_{\nu} \propto \nu^{-2}$ and listed in the \textit{WISE} user manual\footnote{available at this address: http://WISE2.ipac.caltech.edu} for each filter band. When the source was blended in the \textit{WISE} data, we used the \textit{Spitzer}/GLIMPSE data (see Table \ref{glimpse}). GLIMPSE fluxes were directly imported from the GLIMPSE archive. We point out that IGR J14331-6112,  and IGR J17404-3655 are not detected neither in \textit{WISE} or GLIMPSE catalogs whereas IGR J14488-5942 and IGR J17354-3255 are blended both in the GLIMPSE and \textit{WISE} catalogs.

Most of the sources studied here were detected at 3.4 $\mu$m and  4.6 $\mu$m (see Table \ref{WISE}) or at 3.6 and 4.5 $\mu$m (when \textit{Spitzer}/IRAC data were used instead). Two sources (IGR J16318-4848 and IGR J17586-2129) were detected even at 22 $\mu$m which can be explained by a substantial quantity of circumstellar dust, consistent with previous observations of these two sources (see Sections \ref{16318} and \ref{17586}).\\

We also examined the close environment of the sources and searched for cavity-shaped-structures around the sources that might trace an interaction with the close environment. We only detected a cloud with a cavity surrounding IGR J11435-6109 (see Figure \ref{11435_WISE}). This structure may be associated with the HMXB. A more detailed and comprehensive investigation of this aspect will be the topic of a forthcoming paper.

\begin{table*}
\begin{tabular}{llllll}
  \hline
  \hline
  Source name & 3.4 $\mu$m flux (mJy)  & 4.6 $\mu$m flux (mJy) & 12 $\mu$m flux (mJy) & 22 $\mu$m flux (mJy)\\
    \hline 
IGR J10101-5654 & 36.07 $\pm$ 0.764 & 29.14 $\pm$ 0.564 & 8.788 $\pm$ 0.364 & $<$5.227 \\
IGR J11187-5438 & 0.437 $\pm$ 0.024  & 0.239 $\pm$ 0.018 & $<$0.877& $<$5.227\\
IGR J11435-6109 & 8.416 $\pm$ 0.217 & 6.506 $\pm$ 0.174 & 1.576 $\pm$ 0.128 & $<$5.227\\
IGR J16195-4945 & 23.07 $\pm$ 0.638 & 17.19 $\pm$ 0.490 & $<$0.877 & $<$5.227\\
IGR J16318-4848 & 1001 $\pm$ 47 & 1135 $\pm$ 25 & 310 $\pm$ 5.4 & 297 $\pm$ 7.7\\ 
IGR J17586-2129 & 316 $\pm$ 9.3  & 376 $\pm$ 7.3 & 116 $\pm$ 1.8  & 46.97 $\pm$ 2.163 \\
   \hline
\end{tabular}
  \caption{\textit{WISE} counterpart magnitudes.}
  \label{WISE}
\end{table*}

\begin{table*}
\begin{tabular}{lllll}
  \hline
  \hline
  Source name & 3.6 $\mu$m flux (mJy)  & 4.5 $\mu$m flux (mJy) & 5.8 $\mu$m flux (mJy) & 8.0 $\mu$m flux (mJy) \\
    \hline
 IGR J13020-6359 & 11.57 $\pm$ 0.673  & 9.157 $\pm$ 1.025 & 7.269 $\pm$ 0.559 & 4.690 $\pm$ 0.455 \\
 IGR J16320-4751 & 42.04 $\pm$ 1.814 & 42.89 $\pm$ 2.010 & 36.20 $\pm$ 1.553 & 17.76 $\pm$ 0.951 \\
 IGR J16328-4726 & 23.14 $\pm$ 1.052 & 21.40 $\pm$ 1.140 & 16.22 $\pm$ 0.754  & 10.01 $\pm$ 0.405 \\
 IGR J16418-4532 & 11.81 $\pm$ 0.642 & 9.352 $\pm$ 0.566 & 5.855 $\pm$ 0.463  & 3.394 $\pm$ 0.218 \\ 
   \hline
\end{tabular}
  \caption{\textit{Spitzer}/GLIMPSE counterpart magnitudes for the blended or undetected sources in the \textit{WISE} point source catalog.}
  \label{glimpse}
\end{table*}

\section{Discussion}\label{discuss}

%\subsection{H, He and metallic lines as probes of different circumbinary components}

%\subsection{H band spectrum: Brackett series and disk inclination}

In this section we discuss the number of BeHMXB versus the number of sgHMXB currently identified and the implications of this ratio. We focus only on the HMXB for which a spectral type has been derived.\\
Of the 114 HMXB candidates listed in the Liu catalog of HMXB \citep{Liu2006}, 39 are confirmed as BeHMXB and 18 as sgHMXB \citep[see][]{Liu2006}. We therefore considered all \textit{INTEGRAL} source HMXB candidates contained in the Liu catalog and not confirmed yet. For these objects, we retrieved the identification (when available) in the \textit{INTEGRAL} source catalog maintained by J. Rodriguez and A. Bodaghee\footnote{http://irfu.cea.fr/Sap/IGR-Sources}. We added to these sources all the other \textit{INTEGRAL} sources identified since 2006 and the 13 sources identified in the present article (see Table \ref{identif}). This left us with 44 confirmed BeHMXB and 37 confirmed sgHMXB. 
Six sources are mentioned to have peculiar features in the Liu catalog: CI Cam, IGR J16318-4848, Cyg X-3, LS 5039, V4641 Sgr, and SS 433. We add two objects to this list: IGR J10101-5654 and IGR J19113+1533 which seem to show the B[e] phenomenon (see Section \ref{10101_biblio} for IGR J10101-5654 and \citet{Masetti2010} for IGR J19113+1533).\\

\begin{figure*}
\begin{minipage}[t]{1\textwidth}
\begin{center}
\includegraphics[scale=0.5]{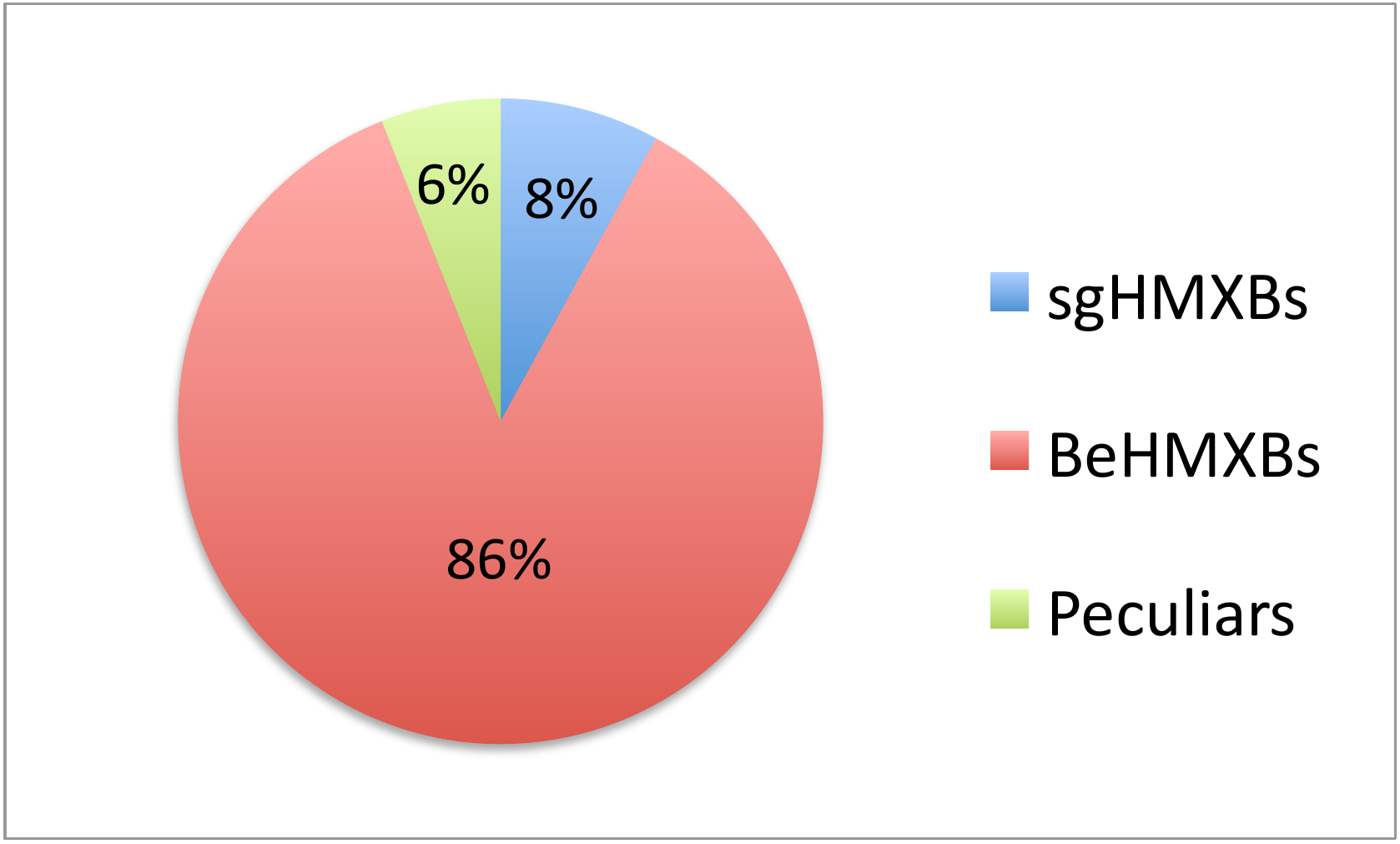}
\includegraphics[scale=0.5]{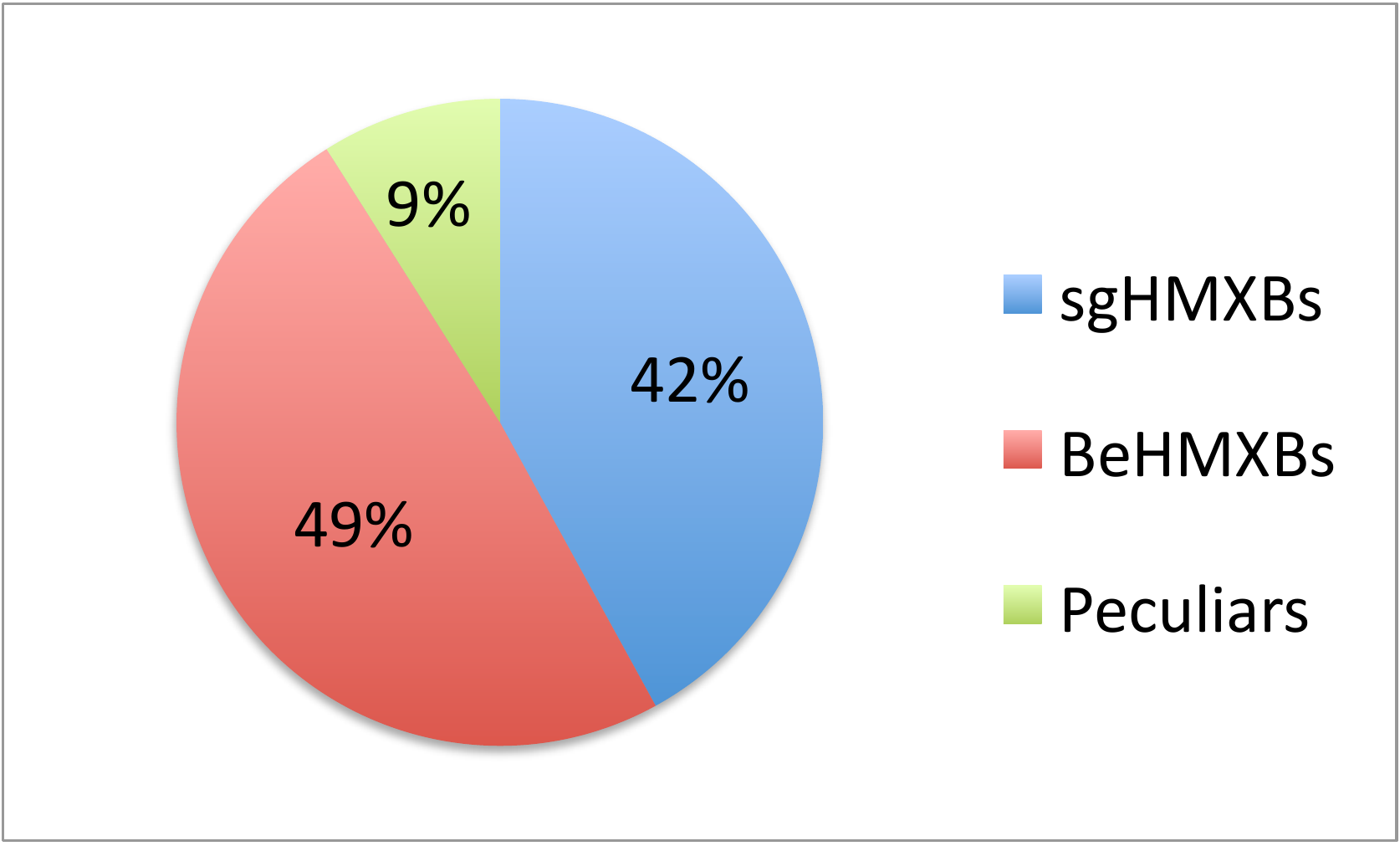}\\
\end{center}
\end{minipage}
\caption{Left panel: Fraction (in percents) of confirmed (with spectral type) sgHMXB, BeHMXB and peculiar HMXB (such as B[e]HMXB) in \citet{Liu2000} catalog. Right panel: Fraction (in percents) of confirmed (with spectral type) sgHMXB, BeHMXB and peculiar HMXB (such as B[e]HMXB) to date.}
\label{stat}
\end{figure*}

Before \textit{INTEGRAL}, there were 42\% confirmed BeHMXB and only 4\% confirmed sgHMXB out of the total number of HMXB catalogd in \citet{Liu2000}. The \textit{INTEGRAL} observations have led to the discovery of a significant number of sgHMXB, causing a dramatic increase in the fraction of HMXB that are sgHMXB. Indeed, based on this study, we  reach the following statistics for the HMXB confirmed by spectral type: 49 \% BeHMXB, 42\% sgHMXB and 9 \% peculiar HMXB (see Figure \ref{stat}). 

 The majority of sgHMXB currently discovered suffers from huge X-ray absorption that explains why these sources were not detected before the \textit{INTEGRAL} launch. Because of this substantial absorption they were not detected in the soft X-ray domain, which was well covered by the previous high-energy instruments. By observing the sky beyond 10 keV, where the photoelectric absorption becomes negligible, \textit{INTEGRAL} enabled to discover these obscured sources, making their population rapidly growing over the last ten years. Moreover, the high sensitivity of the \textit{INTEGRAL} detectors, which enables one to discover sources with a very low quiescent level ($\sim10^{33}$ erg~s\ensuremath{^{-1}}), and their large field of view also facilitated detecting new obscured sgHMXB.

%Moreover, \textit{INTEGRAL} discovered several objects showing peculiar features that challenge our knowledge of the HMXB evolution and formation.

Finally, the number of HMXB discovered with \textit{INTEGRAL} and identified thanks to infrared observations is now high enough to carry out a population study of this family of high-energy sources and to accurately study their formation and evolution. 

%\begin{figure}[th!]
%\begin{center}
%\includegraphics[scale=0.4]{stat.pdf}
%\caption{Fraction (in percents) of confirmed (with spectral type) sgHMXB, BeHMXB and peculiar HMXB (such as B[e]HMXB) to date. }
%\label{stat}
%\end{center}
%\end{figure}

We now would like to place these source-type statistics into the context of HXMB evolution.
HMXB are divided into two classes: sgHMXB and BeHMXB. In both classes, the X-ray source is generally a neutron star powered by accretion of matter coming from the companion star. Accretion in sgHMXB can be driven by Roche lobe overflow or a stellar wind, whereas in BeHMXB, the companion star is not filling its Roche-lobe and the neutron star generally accretes matter when passing through the decretion disk located around the companion. Formation of BeHMXB and sgHMXB is explained by two different scenarii, principally controlled by physical parameters such as the transfer of mass and angular momentum, the spin and metallicity of the stars, or the kick experienced during the supernova event. BeHMXB are formed following the rejuvenation model of \citet{Rappaport1982}, in which the mass transfer engenders the formation of a decretion disk around the companion star. These systems, showing wide orbital eccentricities, are the result of massive binaries that undergo a semi-conservative mass transfer evolution. On the other hand, sgHMXB evolution is driven by the common-envelope phenomenon, during which the compact object enters inside the envelope of the companion star. Thus, its orbital period is rapidly decreasing via a significant loss of orbital angular momentum \citep[see e.g.][]{Chaty2011}.\\%BeHMXB spectral types observed in the Galaxy and Large/Small Magellanic Clouds (respectively  LMC and SMC), between O9 and B2V \citep{McBride2008}, are explained by the fact that stars less massive than 8 M$_{\sun}$  are ejected when the supernova explodes and stars more massive than 22  M$_{\sun}$ become supergiant stars. On the other hand, sgHMXB evolution is driven by the common envelope phenomenon during which the compact object enters inside the envelope of the companion star. Thus, its orbital period is rapidly decreasing via a significant loss of orbital angular momentum \citep[see e.g.][]{Chaty2011}.\\
%Moreover, \citet{McBride2008} show that the spectral distribution of BeHMXB observed in the SMC is consistent with that in the Milky Way despite the lower metallicity environment of the SMC, underlining mainly the key role of loss of angular momentum in the BeHMXB evolution. However, \citet{Linden2010}, investigating young ($<$ 20 Myr) and bright HMXB, find the population to be strongly metallicity dependent, primarily affecting the number of systems which evolve through Roche Lobe overflow or through wind accretion pathway.\\
%When plotting the neutron star spin period P$_{\mathrm{spin}}$  as a function of the orbital period P$_{\mathrm{orb}}$ of the system in the Corbet diagram \citep{Corbet1986}, one realizes that HMXB are localized in distinct places according to their nature (BeHMXB or sgHMXB). A clear correlation between P$_{\mathrm{spin}}$ and P$_{\mathrm{orb}}$ is detected for BeHMXB \citep{Corbet1984}, due to the efficient transfer of angular momentum during accretion phases which is not the case in supergiant systems \citep[see e.g.][Figure 1]{Chaty2011}. Furthermore, \citet{Knigge2011} have investigated the Corbet diagram of BeHMXB and revealed the presence of two sub-populations with different spin and orbital periods and eccentricities. They argue that these two families are probably linked to the two types of supernova that produce neutron stars: iron-core collapse and electron-capture supernova. To improve these studies, it is now important to keep identifying new HMXB in the Milky Way, along with their spin and orbital parameters.\\
\indent\textit{INTEGRAL} observations have raised new questions about HMXB evolution, particularly by discovering two new classes of sgHMXB that were previously undetected. The first one is composed of very obscured sgHMXB, exhibiting huge local extinction, and the others are sgHMXB showing fast and transient outbursts, called SFXTs \citep{Negueruela2006,Negueruela2008}. % \citet{Negueruela2008} explain the X-ray lightcurves of SFXTs by a stellar wind clumpy model, invoking two sub-regions into the stellar wind, differing by the clump density. In this model, the transition between SFXTs and standard sgHMXB can be explained by a different neutron star orbital radius, making the compact object either accreting material from a low density clumpy stellar wind (when the orbital radius is higher than $\sim$ 2 stellar radii) or accreting quasi-continuous wind (when the orbital radius is lower than $\sim$ 2 stellar radii). Other models can explain the SFXT sources, such as the formation of transient accretion discs \citep{Ruffert1996,Ducci2010} or the accretion with centrifugal/magnetic barriers \citep{Bozzo2008}. Likewise, \citet{Lutovinov2013} studied a well defined sample of persistent HMXB in the Milky Way and concluded that SFXT behavior is likely caused by magnetic inhibition of the accretion onto the compact object. \\
Obscured HMXB are exhibiting complex and stratified circumstellar material. Some of these sgHMXB present the B[e] phenomenon (such as IGR J10101-5654 or IGR J16318-4848, see Sections \ref{16318} and \ref{10101_biblio}). \citet{Lamers1998} defined these sources as stars that \textit{i.} emit forbidden lines of [\ion{Fe}{ii}] and [\ion{O}{i}]; \textit{ii.} exhibit a strong mIR excess due to hot circumstellar dust; \textit{iii.} present P-Cygni line profiles, which are characteristic of huge mass-loss. The peculiarities of these objects may be explained by ejections of matter during phases of interaction with the compact object \citep[see e.g.][]{Kraus2010,Wheelwright2012}. Indeed, according to \citet{Wheelwright2012}, at least 50\% of the sgB[e] stars observed at high spatial resolution may be interacting binaries, which raises the possibility that binarity plays an important role in this process. Observing these obscured sgB[e] HMXB sources, which seem to be composed of a dusty disk (see e.g. \citealt{Chaty2012} and Servillat et al. in prep.), is a step forward in the understanding of circumstellar environment of evolved massive stars. Other sgB[e], at a different evolutionary stage, also exhibit a disk probably formed after an efficient mass-transfer \citep[see e.g.][]{Millour2011} and the link with these objects should be investigated.\\
%Two SFXTs are misplaced in the Corbet diagram \citep{Liu2011}: IGR J11215-5952 \citep{Negueruela2005} and IGR J18483-0311 \citep{Rahoui2008}. \citet{Liu2011} suggest that they cannot have evolved from normal main-sequence OB-type stars but rather from O-type emission line stars. This result may reveal the presence of two types of sgHMXB: one directly evolved from normal OB main-sequence star and another evolved from BeHMXB after having suffered a first accretion period.\\
\indent All these results about sgHMXBs can be combined to explain the observed differences by different orbital configurations \citep[see e.g.][]{Chaty2011}: obscured sgHMXB are composed of a neutron star orbiting a supergiant star at a few stellar radii that emits persistent X-ray emission, whereas in SFXTs, a neutron star is orbiting the supergiant in a large and eccentric orbit, accreting clumps of matter when it comes close to the companion star. This scenario describes the properties currently observed in HMXB quite well, even though some questions still remain open. For example, no purely wind accreting sgHMXB with a black hole  has been detected yet (Cygnus X-1 is accreting both wind and disk matter), probably because massive stars lose too much mass to finish their life as a black hole. Likewise, the rarity of HMXB that host a companion Wolf-Rayet star (only Cygnus X-3 is known) was underlined over the years and places constraints on evolution of HMXB \citep[see][]{Linden2012,Lommen2005}. \citet{Lommen2005} also suggested that several X-ray binaries with 
core helium-burning companions should be observed in the Milky Way. They are expected to exhibit high intrinsic absorption mainly due to X-ray absorption by stellar wind. Companion stars of these obscured sgHMXB could evolve as follows \citep{Kogure2007}: O $\rightarrow$ Of  $\rightarrow$  WN (hydrogen-rich) $\rightarrow$ LBV $\rightarrow$ WN (hydrogen-poor) $\rightarrow$ WC $\rightarrow$ SN: starting their life as an O star on the main-sequence, then evolving as an LBV star, and finally finishing their life as a Wolf-Rayet star. sgB[e] and LBV companions are probably at the interface between hydrogen- and helium-burning core and could then partly explain the expected population described by \citet{Lommen2005}.\\
Finally, studying HMXB is of interest for a wide range of astrophysics subjects. Indeed, it has been shown by \citet{Mirabel2011} that black hole HMXB could have determined the early thermal history of the Universe and contributed to its reionization, between 380 000 and 1 billion years after the Big Bang. Moreover, HMXB act as a good tracer of recent star formation in galaxies because of their short lives \citep[see e.g.][]{Mineo2011,Bodaghee2012,Coleiro2013}. It has been shown by \citet{Sana2012} that almost three quarters of the very bright high-mass stars do not live alone, most of them experience interactions such as mass transfer during their life, which makes them an interesting bridge between HMXB and stellar evolution investigations.

\section{Conclusion}

We have presented results of a large nIR observing campaign carried out in 2008 and 2010 using ESO/NTT facilities. These photometric and spectroscopic data allowed us to constrain and/or confirm the spectral type of 13 HMXB in the Milky Way reported in Table \ref{identif}. The proportion of confirmed sgHMXB is now 42\%, 49\% are confirmed BeHMXB, and 9\% are peculiar sources. Moreover, we conducted a systematic search of mIR counterparts in the \textit{WISE} data release in order to detect any mIR excess due to  hypothetical circumstellar dust, but also with the aim of examining the larger-scale environment that presents possible interaction and feedback. Finally, we recommend additional investigations of the three sources presenting a mIR excess (IGR J10101-5654, IGR J11435-6109, and IGR J17586-2129) to assess the physical conditions in their close environment.

\begin{center}
\begin{table*}\begin{tabular}{lccll}
\hline
  \hline
  Source & nIR Counterpart RA (J2000) & nIR Counterpart Dec (J2000) & Previous identification & Derived SpT (this paper) \\
  \hline
  
  IGR J10101-5654 & 10:10:11.87 & -56:55:32.1 & HMXB ? (1,2)  &sgB[e]\\
  IGR J11187-5438 & 11:18:21.21 & -54:37:28.6 & ?  & \textbf{LMXB ?}\\
  IGR J11435-6109 & 11:44:00.30 & -61:07:36.5 &B2III or B0V (3) & B0.5Ve\\
  IGR J13020-6359 & 13:01:58.7 & -63:58:09.0 & Be (4) & B0.5Ve\\
  IGR J14331-6112 & 14:33:08.33 & -61:15:39.7 & BIII or BV, Be ? (5) & Be ?\\
  IGR J14488-5942  & 14:48:43.33  & -59:42:16.3 & Be ? (6) & Oe/Be\\
  IGR J16195-4945 & 16:19:32.20  & -49:44:30.5 & O/B supergiant (7) & ON9.7Iab\\
  IGR J16318-4848 & 16:31:48.31 & -48:49:00.7 & sgB[e] (8) & sgB[e]\\
  IGR J16320-4751 & 16:32:01.75 & -47:52:28.9 & O supergiant ? (9,10) & BN0.5Ia\\
  IGR J16328-4726 & 16:32:37.91 & -47:23:40.9 &O/B supergiant ? (11) & O8Iafpe\\
  IGR J16418- 4532 & 16:41:50.79 & -45:32:25.3 & O/B supergiant (12) & BN0.5Ia\\
  IGR J17200-3116 & 17:20:06.1 & -31:17:02.0 & HMXB ? (2)& ?\\
  IGR J17354-3255  & 17:35:27.60 & -32:55:54.40 & sgHMXB ? (13)& O9Iab \\ 
  IGR J17404-3655 & 17:40:26.85  & -36:55:37.6 & LMXB ? / HMXB ? (3,14) & Be ?\\
  IGR J17586-2129 & 17:58:34.56 & -21:23:21.53 &  HMXB ? (14) & supergiant O/B\\
  \hline
\end{tabular}
  \caption{Summary of results on HMXB spectral type (SpT) presented in this study. Reference list for previous classification: (1) \citet{Tomsick2008}, (2) \citet{Masetti2006}, (3) \citet{Masetti2009}, (4) \citet{Chernyakova2005}, (5) \citet{Masetti2008}, (6) \citet{Corbet2010_1}, (7) \citet{Tomsick2006}, (8) \citet{Filliatre2004}, (9) \citet{Corbet2005_1}, (10) \citet{Chaty2008}, (11) \citet{Fiocchi2013}, (12) \citet{Sguera2006}, (13) \citet{Sguera2011}, (14) \citet{Tomsick2009}.
   }
    \label{identif}
\end{table*}
\end{center}

\begin{acknowledgements}
This work was supported by the Centre National d'Etudes Spatiales (CNES), based on observations obtained with MINE -- the Multi-wavelength \textit{\textit{INTEGRAL}} NEtwork--. JAT acknowledges partial support from NASA through {\em Chandra} Award Number GO1-12046X issued by the {\em Chandra} X-ray Observatory Center, which is operated by the Smithsonian Astrophysical Observatory under NASA contract NAS8-03060. This research has made use of the IGR Sources page maintained by J. Rodriguez \& A. Bodaghee (http://irfu.cea.fr/Sap/IGR-Sources); of data products from the Two Micron All Sky Survey, which is a joint project of the University of Massachusetts and the Infrared Processing and Analysis Center/California Institute of Technology, funded by the National Aeronautics and Space Administration and the National Science Foundation; of the SIMBAD database and the VizieR catalog access tool, operated at CDS, Strasbourg, France; of NASA's Astrophysics Data System Bibliographic Services as well as of the NASA/ IPAC Infrared Science Archive, which is operated by the Jet Propulsion Laboratory, California Institute of Technology, under contract with the National Aeronautics and Space Administration.

\end{acknowledgements}

\bibliographystyle{aa}
\bibliography{biblio}

\end{document}